\documentclass[12pt,preprint]{aastex}
\usepackage{graphicx}
\shortauthors{Allers et. al.}
\shorttitle{Discovery of SDSS~J2249+0044AB}

\begin{document}

\newcommand{\Ks}{\mbox{$K_S$}}
\newcommand{\Lp}{\mbox{$L^{\prime}$}}
\newcommand{\mbol}{\mbox{$m_{\rm{bol}}$}}
\newcommand{\sciencebin}{\hbox{SDSS~J2249+0044AB}}
\newcommand{\sciencebina}{\hbox{SDSS~J2249+0044A}}
\newcommand{\sciencebinb}{\hbox{SDSS~J2249+0044B}}
\newcommand{\degs}{\mbox{$^{\circ}$}}
\newcommand{\perpix}{\mbox{pixel$^{-1}$}}
\newcommand{\etal}{et al.}
\newcommand{\eg}{e.g.}
\newcommand{\ie}{i.e.}

\title{Discovery of a Young L Dwarf Binary, SDSS~J224953.47+004404.6AB\footnote{Some of the data presented herein were obtained at the W.M. Keck Observatory, which is operated as a scientific partnership among the California Institute of Technology, the University of California and the National Aeronautics and Space Administration. The Observatory was made possible by the generous financial support of the W.M. Keck Foundation.}}

\author{K.~N.~Allers\altaffilmark{2,3}}
\affil{Department of Physics and Astronomy, Bucknell University, Lewisburg, PA 17837, USA; k.allers@bucknell.edu}
\author{Michael~C.~Liu\altaffilmark{4}, Trent~J.~Dupuy, Michael~C.~Cushing\altaffilmark{3}} 
\affil{Institute for Astronomy, University of Hawaii, 
2680 Woodlawn Drive, Honolulu, HI 96822, USA}

\altaffiltext{2}{Institute for Astronomy, University of Hawaii, 2680 Woodlawn Drive, Honolulu, HI 96822}
\altaffiltext{3}{Visiting Astronomer at the Infrared Telescope Facility, which is operated by the University of Hawaii under Cooperative Agreement no. NCC 5-538 with the National Aeronautics and Space Administration, Office of Space Science, Planetary Astronomy Program.}
\altaffiltext{4}{Alfred P. Sloan Research Fellow}
%\altaffiltext{1}{Visiting Astronomer at the Infrared Telescope Facility, which is operated by the University of Hawaii under Cooperative Agreement no. NCC 5-538 with the National Aeronautics and Space Administration, Office of Space Science, Planetary Astronomy Program.}
%\altaffiltext{2}{Alfred P. Sloan Research Fellow}
\begin{abstract}
We report discovery of a young 0$\farcs$32 L dwarf binary, \sciencebin, found as the result of a Keck laser guide star adaptive optics imaging survey of young field brown dwarfs.  
Weak \ion{K}{1}, \ion{Na}{1}, and FeH features as well as strong VO absorption in the integrated-light $J$-band spectrum indicate a low surface gravity and hence young age for the system.  From spatially resolved $K$-band spectra we determine spectral types of L3 $\pm$ 0.5 and L5 $\pm$ 1 for components A and B, respectively.  
\sciencebina\ is spectrally very similar to G196-3B, an L3 companion to a young M2.5 field dwarf.  Thus, we adopt 100~Myr (the age estimate of the G196-3 system) as the age of \sciencebin, but ages of 12--790~Myr are possible.  By comparing our photometry to the absolute magnitudes of G196-3B, we estimate a distance to \sciencebin\ of 54 $\pm$ 16~pc and infer a projected separation of 17 $\pm$ 5~AU for the binary.  Comparison of the luminosities to evolutionary models at an age of 100~Myr yields masses of 0.029 $\pm$ 0.006 and 0.022$^{+0.006}_{-0.009} M_\odot$ for \sciencebina\ and B, respectively.  Over the possible ages of the system (12--790~Myr), the mass of \sciencebina\ could range from 0.011 to 0.070~$M_\odot$ and the mass of \sciencebinb\ could range from 0.009 to 0.065~$M_\odot$.  Evolutionary models predict that either component could be burning deuterium, which could result in a mass ratio as low as 0.4, or alternatively, a reversal in the luminosities of the binary.  We find a likely proper motion companion, GSC~00568-01752, which lies 48$\farcs$9 away (a projected separation of 2600~AU) and has Sloan Digital Sky Survey and Two Micron All Sky Survey colors consistent with an early M dwarf.  We calculate a photometric distance to GSC~00568-01752 of 53 $\pm$ 15~pc, in good agreement with our distance estimate for \sciencebin.  The space motion of \sciencebin\ shows no obvious coincidence with known young moving groups, though radial velocity and parallax measurements are necessary to refine our analysis.  The unusually red near-IR colors, young age, and low masses of the binary make it an important template for studying planetary-mass objects found by direct imaging surveys.

\end{abstract}


\section{Introduction}

\setcounter{footnote}{4}

Brown dwarfs are objects with masses too low to sustain hydrogen burning in their cores.  Because they continually cool over their lifetimes, substellar objects are much more luminous when they are young.  For example, a 0.02 $M_\odot$ brown dwarf at 10~Myr is $\sim$5000 times more luminous than it will be at 3~Gyr, roughly the age of the general field population \citep{burrows01}. Similarly, planetary-mass objects ($M \lesssim 0.012 M_\odot$) are expected to be more luminous when young.  In fact, the first directly imaged planets outside of our solar system were detected as companions to young host stars \citep{kalas08, marois08}.  Unfortunately, the fundamental properties (masses and temperatures) of these planets are difficult to determine because their bright host stars inhibit a detailed spectroscopic study.  Photometry of the planets can be compared to evolutionary models \citep{baraffe03,fortney08} to infer masses and temperatures, but such models remain untested in the planetary-mass regime.  
% Thus, young brown dwarfs provide a unique opportunity for direct detection and study of objects with masses similar to extrasolar planets.

Young brown dwarfs provide a unique opportunity for direct study of planetary-mass companions.  Extremely low-mass objects ($\lesssim$1~M$_{Jupiter}$) are within the reach of direct detection because such objects are self-luminous at young ages.  A brown dwarf primary is itself faint, which means the properties of the planetary-mass companion are not masked by a bright parent star.  To date, 2MASSW J1207334-393254 (hereafter 2MASS~1207)
is the only known young brown dwarf having a planetary-mass companion, 2MASS~1207b \citep{chauvin05,song06}. 
2MASS~1207b has proven to be quite interesting, as it is under-luminous for its temperature and age compared to evolutionary models \citep{mohanty07,biller07}.  It is unclear if 2MASS~1207b formed in an unusual way \citep{mamajek07}, is occulted by a circumstellar disk \citep{mohanty07}, or if the problem lies in the evolutionary models themselves.  Finding and characterizing additional planetary companions to young brown dwarfs can provide critical tests for evolutionary models as well as empirical templates for comparison to other young planets.

%First, studies of brown dwarf multiplicity find that brown dwarf binaries tend to be in nearly equal-mass systems.  Thus, lower mass primaries are statistically more likely to harbor planetary-mass companions.  Young brown dwarfs are the lowest mass free-floating objects.  Second, when young, planetary mass companions are within the reach of direct detection, because substellar objects are much more luminous when young.  For example, Jupiter-mass objects are predicted to be bright when 1~Myr old (K-band absolute magnitude of xxx), but by the time they reach the age of the field population (~1 Gyr), they are far to faint (K mags of xxx) for direct detection.  **2M1207 stuff*

%In the past decade, numerous young brown dwarfs have been discovered.  
Most of the known young brown dwarfs are members of star-forming regions \citep[e.g.,][]{muench07,lucas06, allers06} which renders them unsatisfactory for imaging surveys searching for planetary-mass companions for a number of reasons.  First, brown dwarfs in star-forming regions are often attenuated by natal cloud material, which can inhibit the determination of the object's fundamental properties.  Second, even the nearest star-forming regions are quite distant ($\sim$125~pc) making the planetary-mass objects faint.  Lastly, current imaging surveys can reach angular resolutions of $\sim$80 mas, which at a distance of 125~pc corresponds to $\sim$10~AU.  
Field dwarf binaries tend to have small (3--10 AU) separations \citep{burgasser07}, which means if companions to young brown dwarfs have similar separations, they would remain unresolved.
%An imaging survey of nearby, young, {\it field} brown dwarfs is ideal for detecting planetary companions, even at close separations. 

To search for planetary-mass companions at small separations we are carrying out a Keck laser guide star adaptive optics (LGS AO) survey to image young ($\lesssim$100~Myr), nearby ($\lesssim$80~pc) brown dwarfs.  This paper presents the early results of our survey \citep[see also][]{allers09} which is part of a larger, ongoing effort using LGS~AO to study the multiplicity of brown dwarfs and determine their properties \citep[e.g.,][]{liu05,liu06,dupuy09,liu10}.
%While these young brown dwarfs have nominal masses as low as a few M$_{Jupiter}$, the large distance to even the nearest star-forming regions ($\sim$125~pc) means that: only wide separation companions can be detected by current imaging surveys.  Ideally, one would like to study young objects that are nearby so.  

%Young field objects are usually found in three ways: as companions to known young field stars \citep[e.g.,][]{rebolo98}, via association with known moving groups \citep[e.g.,][]{gizis02}, or serendipitously as a part of searches for older field dwarfs \citep[e.g.,][]{kirkpatrick06}. 
Using Keck LGS~AO, we have discovered a new young binary system, SDSS~J224953.47+004404.6AB (hereinafter \sciencebin).
\sciencebin\ was first identified by \citet{geballe02} as an L5 dwarf based on near-IR spectroscopy.  \citet{hawley02} obtained optical spectroscopy of \sciencebin\ and assigned it a spectral type of L3.  
The discrepant near-IR and optical spectral types and the unusually red near-IR colors ($J-K$=2.05 mags) of \sciencebin\ were emphasized by \citet{leggett02} and \citet{knapp04}.  \citet{nakajima04} were the first work to point out the spectral peculiarities of \sciencebin, namely weak \ion{K}{1} absorption lines in its $J$-band spectrum and the prominent triangular shape of its $H$-band spectrum.  Other work \citep[e.g.,][]{lucas01,gorlova03} attributes these spectral peculiarities to a low-gravity (young) atmosphere.

In this paper, we present multi-wavelength imaging and spectroscopy used to determine the properties of \sciencebin.

%Recently, samples of young field objects have been studied in detail \citep{cruz09,kirkpatrick08}, improving our understanding of the spectral signatures of youth and the evolution of substellar objects.

%\sciencebin\ was first identified by \citet{geballe02} as L5 dwarf based on UKIRT/CGS4 and Keck/NIRSPEC spectroscopy.  
%In Table 4, they derive spectral types of L6.5 and L3 based on the H$_2$O 1.5 $\mu$m and CH$_4$ 2.2$\mu$m indices.  They also report SDSS photometry of r*=23.88$\pm$0.57, i*=22.05$\pm$0.24, and z*= 19.47$\pm$0.1. J-K, H-K and K mags of 2.03 $\pm$0.07, 1.04$\pm$0.07, and 14.43$\pm$0.05 (UKIRT--MKO system).  \citet{leggett02} report UKIRT UFTI Z photometry of 18.24$\pm$0.05 and the same MKO JHK photometry as reported by \citet{geballe02}.
%\citet{hawley02} assigned a spectral type of L3$pm$1 based on a composite optical spectrum.  They also report SDSS photometry of 23.98, 22.03 and 19.44 in r*,i*,and z*, with uncertainties of ($>$0.5, $>$0.2, and 0.1).  \citet{golimowski04} report \Lp\ photometry of 12.71 $\pm$0.07. \citet{knapp04} report that \sciencebin\ is variable at the 0.1 mag level, report new MKO JHK photometry and give weighted mean photometry of J=16.47 J-K=2.05, J-H=1.11, and H-K=0.94.

%\citet{nakajima04} used Subaru CISCO to obtain R~400 spectra in JH.  They found that the 1.1 $\mu$m H$_2$O feature is weaker in \sciencebin\ than is seen in field L5 dwarfs, and the \ion{K}{1} lines are weaker than most L5s, the H-band is very peaky.

\section{Observations}

\subsection{Keck LGS~AO/NIRC2 Imaging}

We imaged \sciencebin\ on 2006 October 13 and 2008 September 8 (UT) using the facility IR camera, NIRC2, of the Keck II Telescope on Mauna Kea, Hawaii and the LGS~AO system \citep{wizinowich06, vandam06}. 
 We used the narrow field camera (10$\farcs$2 $\times$ 10$\farcs$2 field of view) and the Mauna Kea Observatories (MKO) $J,H,$\Ks, and \Lp\ filters \citep{simons02,tokunaga02}.  
For the $JH$\Ks\ filters, we used an exposure time of 60~s and obtained a series of six dithered images in each filter, offsetting the telescope by a few arcseconds.  
The images were reduced in the standard fashion (dome flat-fielded, median sky-subtracted, registered and stacked).  
The \Lp-band data were taken using a 10-point dither pattern with 200 coadds of 0.3~s integration per image.  
The \Lp-data were reduced in a similar manner as the $JH$\Ks\ data, except that flat fields were constructed from the science frames themselves and sky subtraction was done in a pairwise fashion using consecutive frames.  

To measure the flux ratios and relative positions of the binary's two components, we used an analytic model of the point-spread function (PSF) as the sum of three elliptical Gaussians \citep{liu06,allers09}.
We used the astrometric calibration from
\citet{2008ApJ...689.1044G}, with a pixel scale of
9.963$\pm$0.005~mas~\perpix\ and an orientation for the detector's
$+y$-axis of $+$0$\fdg$13$\pm$0$\fdg$02 east of north.  After applying
the distortion correction developed by B. Cameron (2007, private
communication), the resulting mean astrometry changed by less than
1$\sigma$; however, the rms errors were significantly improved.  For
example, in the \Ks\ band data set the measured separation scatter was
reduced from 0.7 to 0.2~mas and the position angle (P.A.) scatter from 0$\fdg$14 to
0$\fdg$03.  Table~\ref{tbl:obs} summarizes the resulting astrometry
and binary flux ratios, with uncertainties derived from the rms scatter of
measurements for individual dithers.  For objects at $\gtrsim$0$\farcs$8 separation, the final stacked mosaics reach 10$\sigma$ detection limits for point sources of $\sim$21.8, $\sim$22.5 and $\sim$21.6 mags at $JH$\Ks, respectively.  The separation and P.A. of the system show no significant change between our 2006 October 13 and 2008 September 8 (UT) observations.  Given the proper motion of \sciencebin\ \citep[$\mu = 82.8\pm3.7$~mas~yr$^{-1}$, PA=$85\fdg4\pm2\fdg5$][]{bramich08}, our two epochs of imaging indicate that the system is co-moving and physically associated.
%The proper motion of \sciencebin\ \citep[+89$\pm$10, +18$\pm$8~mas~yr$^{-1}$][]{scholz09} combined with the fact that over 2 epochs of imaging, the separation and position angle of the system do not change significantly, indicate that \sciencebina\ and B are co-moving and thus physically associated.  

In order to assess additional systematic errors in our NIRC2
measurements, we applied our PSF-fitting routine to simulated binary
images.  Only the \Ks\ and \Lp\ data sets had appropriate PSF
reference observations (single objects with similar Strehl and FWHM),
and we found negligible (below 1$\sigma$) systematic offsets in these
simulations.  
%In fact, the \Ks\ band rms errors were somewhat smaller
%in the simulations than determined from our dithered images.  
We also note that astrometry measured in multiple bandpasses in 2006 October
is remarkably consistent: the $\chi^2$ of the separation and
P.A. measurements are 0.60 and 0.28, respectively, for an expected
value of 1.37.  Such good agreement implies that our astrometric
errors are reasonable.  We also found insignificant offsets
($<$~0.01~mas, $<$~0$\fdg$006) due to differential chromatic
refraction (DCR) by using the resolved \Ks\ band spectra of the two
components and template $J$ and $H$ band spectra to derive offsets in
the same manner as \citet{dupuy09}.  Finally, we accounted
for the uncertainty in the astrometric calibration by adding these
errors in quadrature to the separation and P.A. measurements.

\subsection{IRTF/SpeX Near-IR Spectroscopy}

\begin{figure}
\vskip 0.5in
\centerline{\includegraphics[width=6in,angle=90]{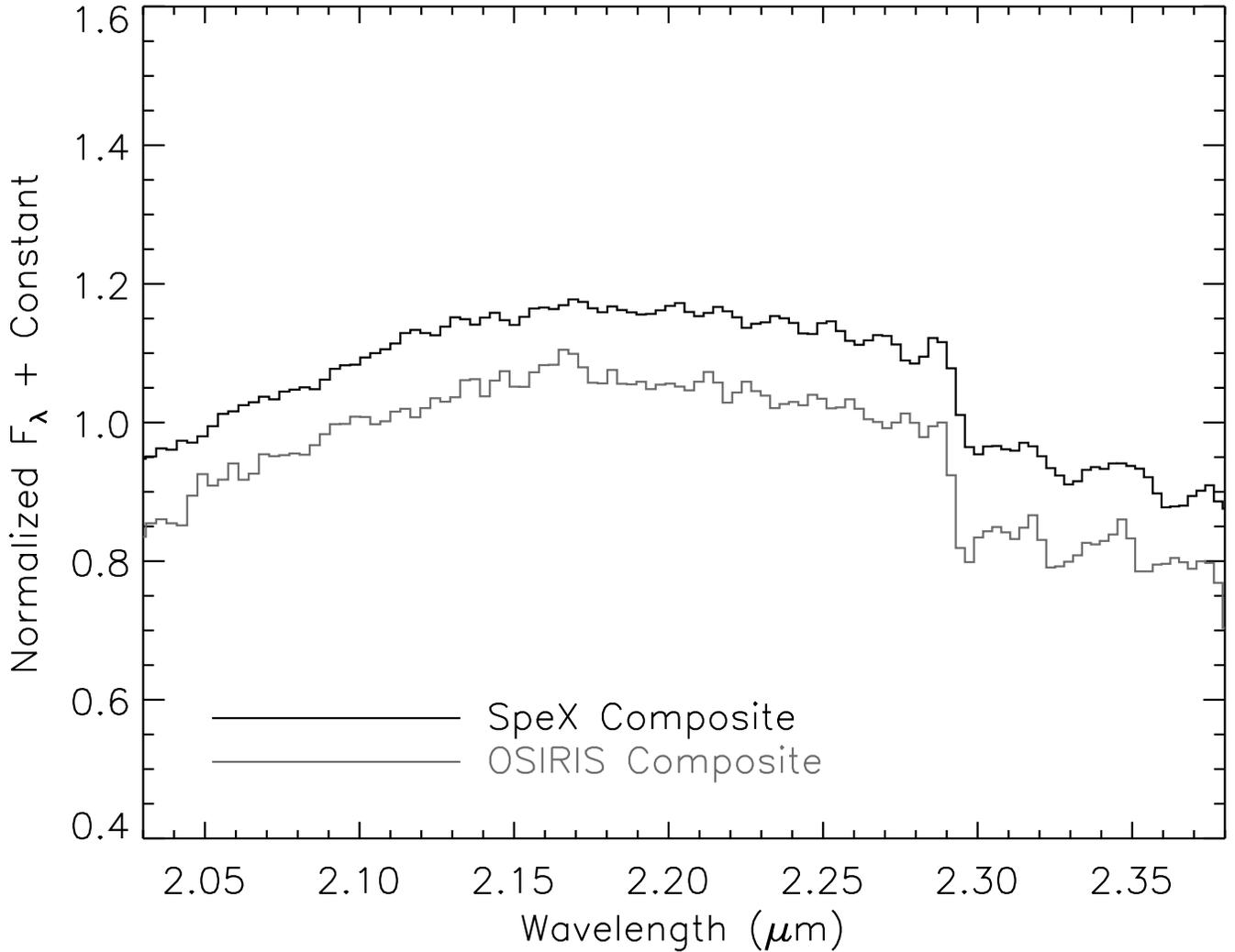}}
\vskip 2ex
\caption[Comparison of Composite $K$-band Spectra]
{\label{spex} Composite K-band spectra of \sciencebin.  Both spectra have been normalized by their medians, and the SpeX integrated-light spectrum (black) has been offset by 0.1.  The Keck~II/OSIRIS composite spectrum (gray) was created by flux calibrating the individual spectra of \sciencebina\ and B using the magnitudes in Table \ref{binphot}, adding the spectra together and smoothing to the SpeX prism resolution (R$\sim$150).  There is good agreement between the two spectra, indicating that our spectral extraction and telluric correction yield consistent results.}
\end{figure}

\begin{figure}
\vskip 0.5in
\centerline{\includegraphics[width=6in,angle=90]{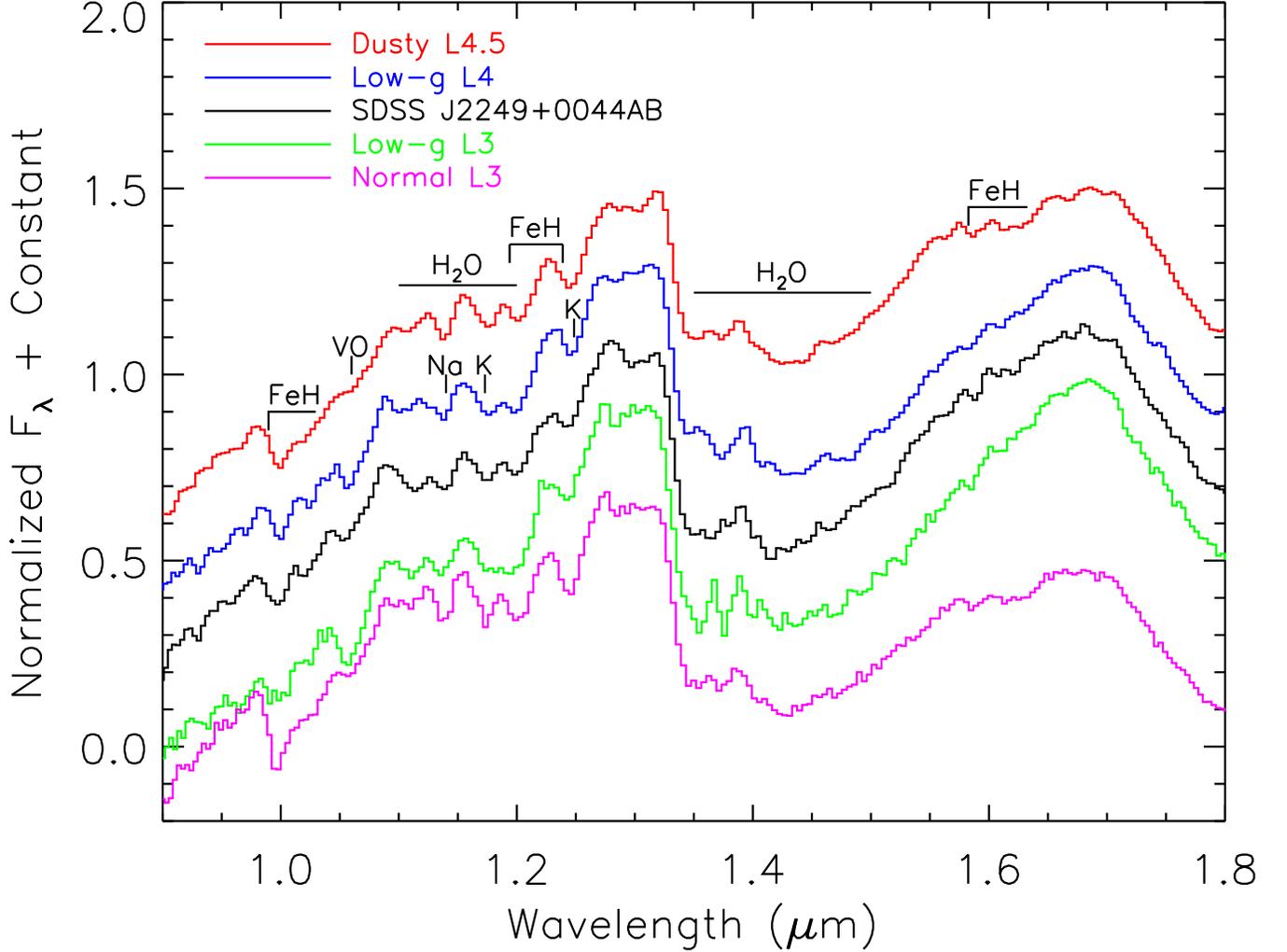}}
\vskip 2ex
\caption[Comparison of J and H band Spectra]
{\label{compnir} Composite near-IR spectrum of \sciencebin\ (black), compared to an L4.5 dwarf with a dusty photosphere (2MASS~2224-0158, red), a low-gravity L4 dwarf (2MASS~0501-0010, blue), a low-gravity L3 dwarf (G196-3B, green), and a normal field L3 dwarf \citep[2MASS~1506+1321;][magenta]{burgasser07}.  The spectra are median-normalized and offset by constants (in intervals of 0.2).  The FeH, \ion{K}{1}, and \ion{Na}{1} absorption features in the spectrum of \sciencebin\ are weaker and the VO absorption is stronger than seen in the normal and dusty L dwarf spectra, indicating a low-gravity (and hence young age) for \sciencebin.}
\end{figure}

Integrated-light near-IR spectroscopy of \sciencebin\ was obtained on 2006 November 19 (UT) using the SpeX spectrograph \citep{rayner03} on the NASA Infrared Telescope Facility (hereinafter IRTF).  
A series of 18 exposures of 180 seconds each were taken, nodding along the slit, for a total integration time of 54 minutes.  
The seeing recorded by the IRTF was 0$\farcs$5.  
The data were taken using the Low-Res prism with the 0$\farcs$5 slit aligned with the parallactic angle, producing a 0.8--2.5~$\mu$m spectrum with a resolution (R=$\lambda/\Delta\lambda$) of $\sim$150.  
For telluric correction of our \sciencebin\ spectrum, we observed a nearby A0V star, HD~216807 and obtained calibration frames (flats and arcs) in between sets of \sciencebin\ observations. 
We flux-calibrated our spectrum of \sciencebin\ using integrated-light $JHK$ photometry from \citet{knapp04} and published Vega flux densities for the MKO filter system \citep{tokunaga05}.
For comparison to our \sciencebin\ spectrum, we obtained SpeX spectra of 2MASS~J05012406-0010452 \citep[L4, hereinafter 2MASS~0501-0010;][]{cruz09,reid08}, 2MASS~J22244381-0158521 \citep[L4.5, hereinafter 2MASS~2224-0158;][]{kirkpatrick00} and G196-3B \citep[L3;][]{cruz09} on 2008 September 24, 2008 November 30, and 2009 January 28 respectively.  
We used the same instrument setup as for our spectrum of \sciencebin\ and obtained total integration times of 20--28 minutes per source.  
The spectra were reduced using the facility reduction pipeline, Spextool \citep{cushing04}, which includes a correction for telluric absorption following the method described in \citet{vacca03}.    
Our SpeX spectra are presented in Figures \ref{spex} and \ref{compnir}.

\subsection{Keck LGS~AO/OSIRIS Imaging and Spectroscopy}

On 2008 September 9 (UT), we obtained imaging of \sciencebin\ using Keck LGS~AO and the imaging camera of OSIRIS \citep{larkin03}.  
The field of view of the OSIRIS imager is 20$\farcs$4 $\times$ 20$\farcs$4 with a plate scale of 20 mas~pixel$^{-1}$.  
We obtained a series of seven dithered images using the $Zbb$ ($\lambda_{\rm{center}}$=1.0915~$\mu$m, $\Delta \lambda$=0.2203~$\mu$m) and the $Hn1$ ($\lambda_{\rm{center}}$=1.5037~$\mu$m, $\Delta \lambda$=0.0747~$\mu$m) filters with exposure times of 120~s and 180~s per image, respectively. 
The resulting images were dome-flatfielded, median sky-subtracted, registered and stacked.  

\begin{figure}
\vskip -8in
% --- vertical layout ---
\vskip 7.5in
\hskip 3.1in
\centerline{\includegraphics[width=6in,angle=90]{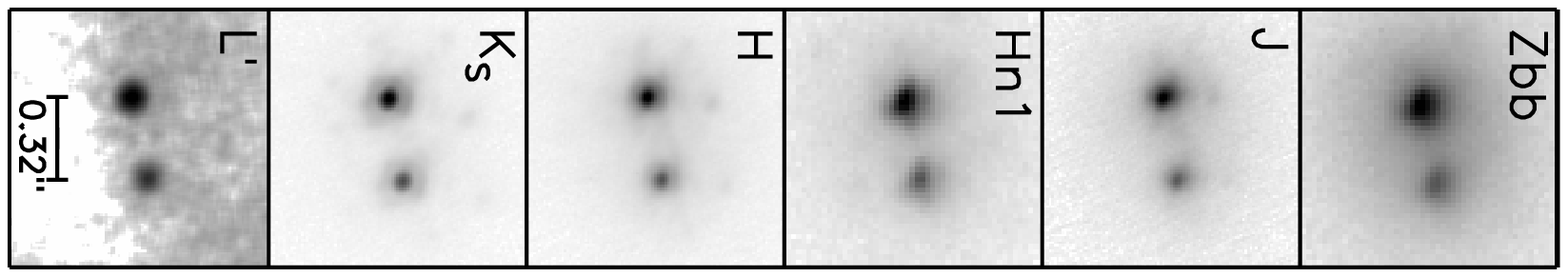}}
\vskip -6.02in
\hskip 4.2in
\centerline{\includegraphics[width=6in,angle=90]{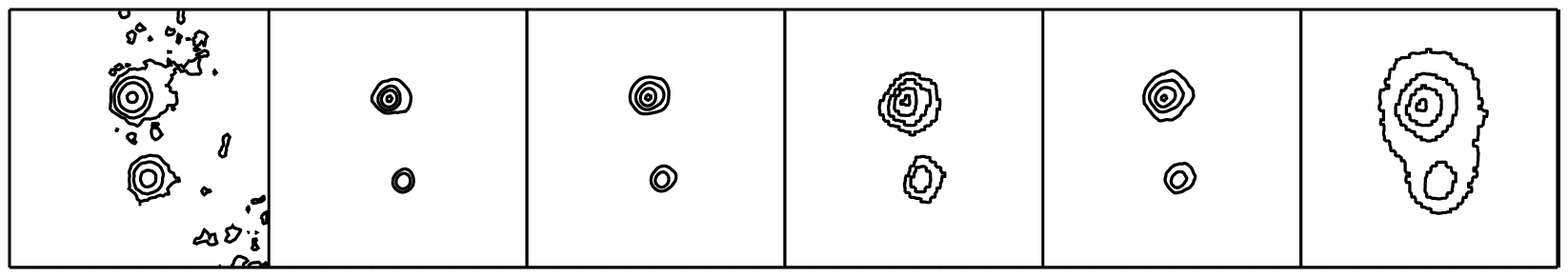}}
\vskip 2ex
\caption{\label{nircims} Imaging of \sciencebin\ from Keck LGS
  AO.  North is up and east is left.  Each image is 0$\farcs$75 on a
  side. The grayscale image uses a square-root intensity stretch.
  Contours are drawn from 90\%, 45\%, 22.5\%, and 11.2\% of
  the peak value in each bandpass.}
\end{figure}

We measured the flux ratios and astrometry of the binary using a similar technique as for our NIRC2 images (Section 2.1). 
For the OSIRIS imager, there is no astrometric calibration available,
so we used the nominal pixel scale of 0$\farcs$020~\perpix.  Comparing
the OSIRIS astrometry from UT 2009 September 9 to the NIRC2
measurements from the previous night, we find a 2.8$\sigma$ difference
in separation, implying a systematic error of 4\% in the OSIRIS imager
pixel scale.  A pixel scale of 0$\farcs$0208~\perpix\ would bring the
two data sets into agreement.  There is no significant P.A. offset
between the two data sets to within 0$\fdg$6.  In the following analysis,
we only require the flux ratios from the OSIRIS imaging, which is not
affected by astrometric calibration of OSIRIS.
A summary of the data is presented in Table \ref{tbl:obs}, and the images are displayed in Figure \ref{nircims}.  
To flux calibrate our OSIRIS photometry, we compute synthetic integrated-light $Zbb$ and $Hn1$ flux densities from our SpeX spectrum of \sciencebin. 
Following the method of \citet{tokunaga05}, we calculate Vega flux densities of 4.87$\times 10^{-9}$ and 1.55 $\times 10^{-9}$ $W~m^{-2}~\mu m^{-1}$ for the $Zbb$ and $Hn1$ filters, respectively.  We then convert the synthetic integrated-light $Zbb$ and $Hn1$ flux densities of \sciencebin\ into Vega-system magnitudes and use our measured flux ratios to calculate the magnitudes of each component (Table \ref{binphot}).

\begin{figure}
\vskip 0.5in
\centerline{\includegraphics[width=6in,angle=90]{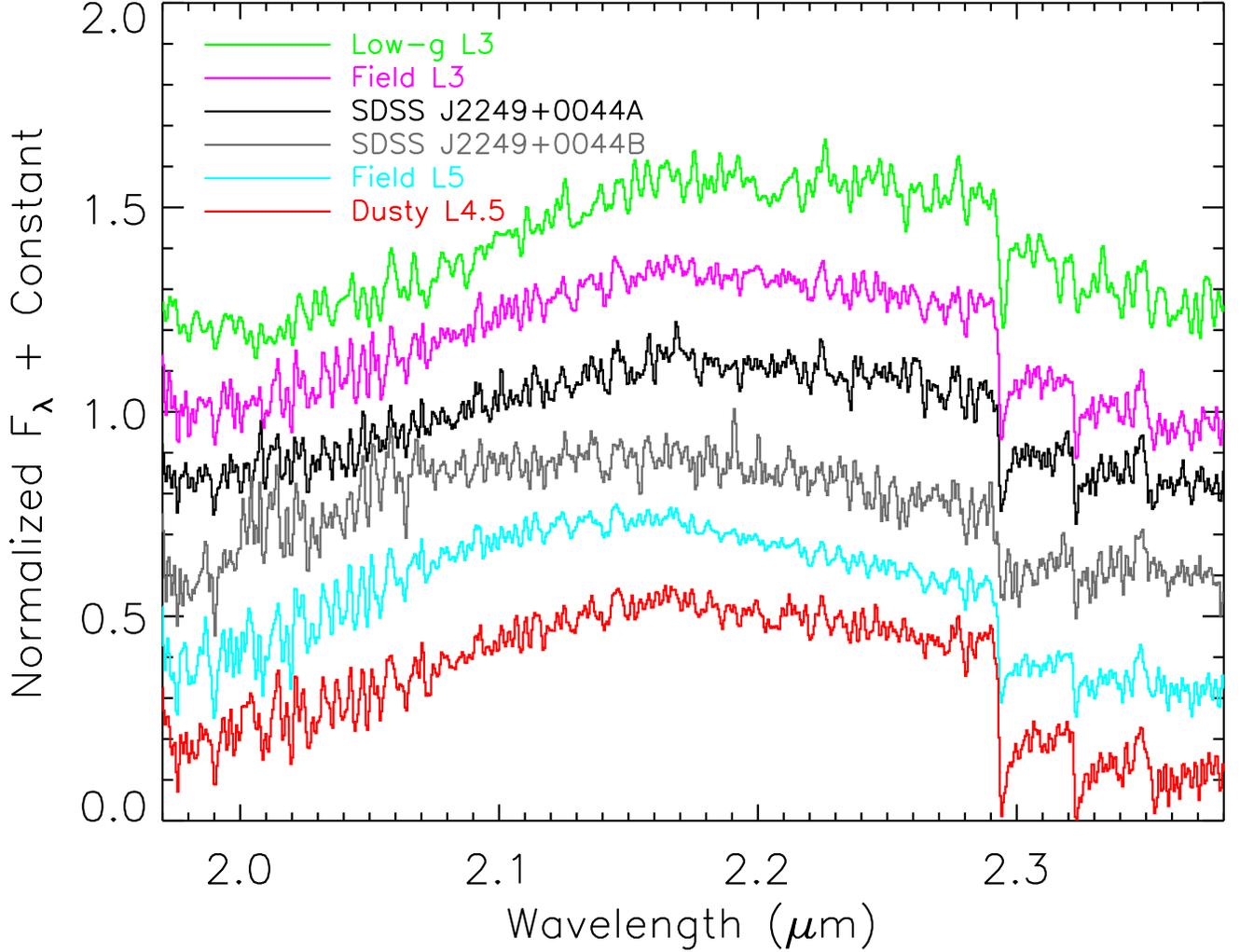}}
\vskip 2ex
\caption[Comparison of K-band Spectra]
{\label{Kspec} Keck~II/OSIRIS $K$-band spectra of \sciencebina\ and B, compared to IRTF/SpeX spectra of a low-gravity L3 dwarf (G196-3B, green), a field L3 dwarf (2MASS~1506+1321, magenta), a field L5 dwarf (2MASS~1507-1627, cyan) and an L4.5 dwarf with a dusty photosphere (2MASS~2224-0158, red).  The spectra of \sciencebina\ and B have been smoothed to the SpeX spectral resolution (R$\sim$2000).  The spectra of field and dusty dwarfs are from \citet{cushing05}, and the G196-3B spectrum is from \citet{allers07}.}
\end{figure}

We obtained spatially resolved $K$-band spectroscopy of \sciencebin\ on 2008 August 23 (UT) using the facility near-IR integral field spectrograph, OSIRIS, and LGS~AO on the Keck-II Telescope.
We selected the 35 mas pixel scale for our observations.  
We observed \sciencebin\ using the $Kbb$ spectrograph filter (1.96--2.38~$\mu$m), taking six dithered exposures of 600 s each, for a total of 1 hr on source integration.   
The FWHM, as measured on the stacked two-dimensional (2D) images of \sciencebin, was 70 mas; thus the binary ($\rho$=322.8 mas) is well resolved. 
Immediately following our observations of \sciencebin, we obtained spectra of a nearby A0V star, HD~210501, in order to correct for telluric absorption.  
The initial reduction from 2D images to 3D data cubes was accomplished using the OSIRIS data reduction pipeline \citep{krabbe04}.  
The individual spectra for each component were then extracted from the 3D data cubes by summing the flux in fixed apertures of 175 $\times$ 175 mas at each wavelength.  
The resulting spectra for each component were median-combined together, and uncertainties were determined from the standard deviation at each wavelength.
Telluric correction and flux calibration were performed using the observations of the A0~V standard and the technique described in \citet{vacca03}.  
The resulting 1.96--2.38~$\mu$m spectra of \sciencebina\ and B (Figure \ref{Kspec}) have a resolution ($\lambda/\Delta\lambda$) of $\sim$3800 and a median signal-to-noise ratio (S/N) per resolution element of $\sim$55 and $\sim$30 for \sciencebina\ and B, respectively.

\subsection{\emph{Spitzer}/IRAC Imaging}

We measured IRAC photometry for \sciencebin\ from observations taken as a part of \emph{Spitzer} program 50059 (PI: A. Burgasser).  Using the IDLPhot package,\footnote{http://idlastro.gsfc.nasa.gov/} we measured aperture photometry from mosaicked images created by the \emph{Spitzer} Science Center (pipeline version 18.7.0).  We calculated the flux in a 3$\farcs$6 aperture, using a background annulus of 12--24\arcsec, and applied the aperture corrections recommended by the IRAC Data Handbook.  Table \ref{binphot} lists our measured IRAC photometry for \sciencebin.
%We then converted the measured flux from MJy~sr$^{-1}$ to Jy~pixel$^{-1}$ (using the pixel size in the image header) and calculated magnitudes using zero magnitude fluxes from the IRAC Data Handbook.  
The uncertainties in our measured photometry (including photometric and calibration uncertainties) were less than 0.02 mag, but we choose to adopt the more conservative 0.05 mag uncertainty recommended by the IRAC Data Handbook.  As a check, we measured photometry for five young L3 and L4 dwarfs with published IRAC photometry \citep{luhman09} and found that our photometry agreed to better than 0.02 mag with previously published results.  

\section{Analysis}

\subsection{Spectral Type}

\indent \sciencebin\ was originally assigned a composite near-IR spectral type of L5$\pm$1 by \citet{geballe02} and an optical spectral type of L3 by \citet{hawley02}.  
%It is important to note, however, that the \citet{geballe02} spectral type is based on their H$_2$O 1.5 $\mu$m and CH$_4$ 2.2$\mu$m indices which yield spectral types of L6.5 and L3 respectively. 
The spectral type sensitive index ($\langle F_{\lambda=1.550-1.560}\rangle/\langle F_{\lambda=1.492-1.502}\rangle$) of \citet{allers07}, which is calibrated for young and field stars with optical spectral types, suggests a spectral type of L4~$\pm$~1, in agreement with both the \citet{geballe02} and \citet{hawley02} spectral types.

We determine spectral types for each component from the spatially resolved OSIRIS $K$-band spectroscopy of \sciencebina\ and B shown in 
Figure \ref{Kspec}.  For comparison we also show spectra of a young L3 dwarf, G196-3B from \citet{allers07}; an L4.5 type dwarf with a dusty photosphere, 2MASS~2224-0158 from \citet{cushing05}; and ``normal'' field L3 and L5 dwarfs, 2MASS~J15065441+1321060 and 2MASS~J15074769-1627386 (hereinafter 2MASS~1506+1321 and 2MASS~1507-1627) from \citet{cushing05}.  
The $K$-band spectra of young field dwarfs do not appear to differ significantly in shape or features from the spectra of field dwarfs.  
Qualitatively, the spectra of \sciencebina\ and B are quite similar to the spectra of field L3 and L5 dwarfs respectively, and \sciencebina\ appears to be of a similar spectral type to G196-3B \citep[young L3,][]{cruz09}.  

Another method of assigning spectral type is using spectral indices.  
Table \ref{indices} lists the index values and derived spectral types for \sciencebina\ and B from $K$-band spectral indices available in the literature.  
These indices, however, are calibrated for field dwarfs, and their applicability to young objects has not been proven.  
We tested the indices in Table \ref{indices} on our spectrum of G196-3B, and found that each index reproduced its L3 optical spectral type to within one subtype.  
We also compared our $K$-band spectra of \sciencebina\ and B to L dwarf spectra from \citet{cushing05} and determined the best fitting L dwarf spectrum via $\chi^2$ minimization.  
The fits to the spectra of \sciencebina\ and B show minima in $\chi^2$ at spectral types of L3$\pm$1 and L6$\pm$2, respectively.\footnote{Uncertainties in the $\chi^2$ fit are determined from the spectral types of \citet{cushing05} spectra where $\chi^2 = \chi^2_{\rm{min}} + 1$.}  
%Due to the lower S/N of the \sciencebinb\ spectrum, $\chi^2$ is roughly flat as a function of spectral type, and we were unable to determine the best-fitting L-dwarf spectrum.  
From the weighted mean of the spectral types determined from $\chi^2$ fitting and calculated from spectral indices (Table \ref{indices}), we assign spectral types of L3$\pm$0.5 and L5$\pm$1 to \sciencebina\ and B, respectively.\footnote{Uncertainties in the spectral types were determined from the weighted average variance of the results of $\chi^2$ fitting  combined with the values reported in Table \ref{indices}.}

\subsection{Age}
\begin{figure}
\vskip 0.5in
\centerline{\includegraphics[width=4in]{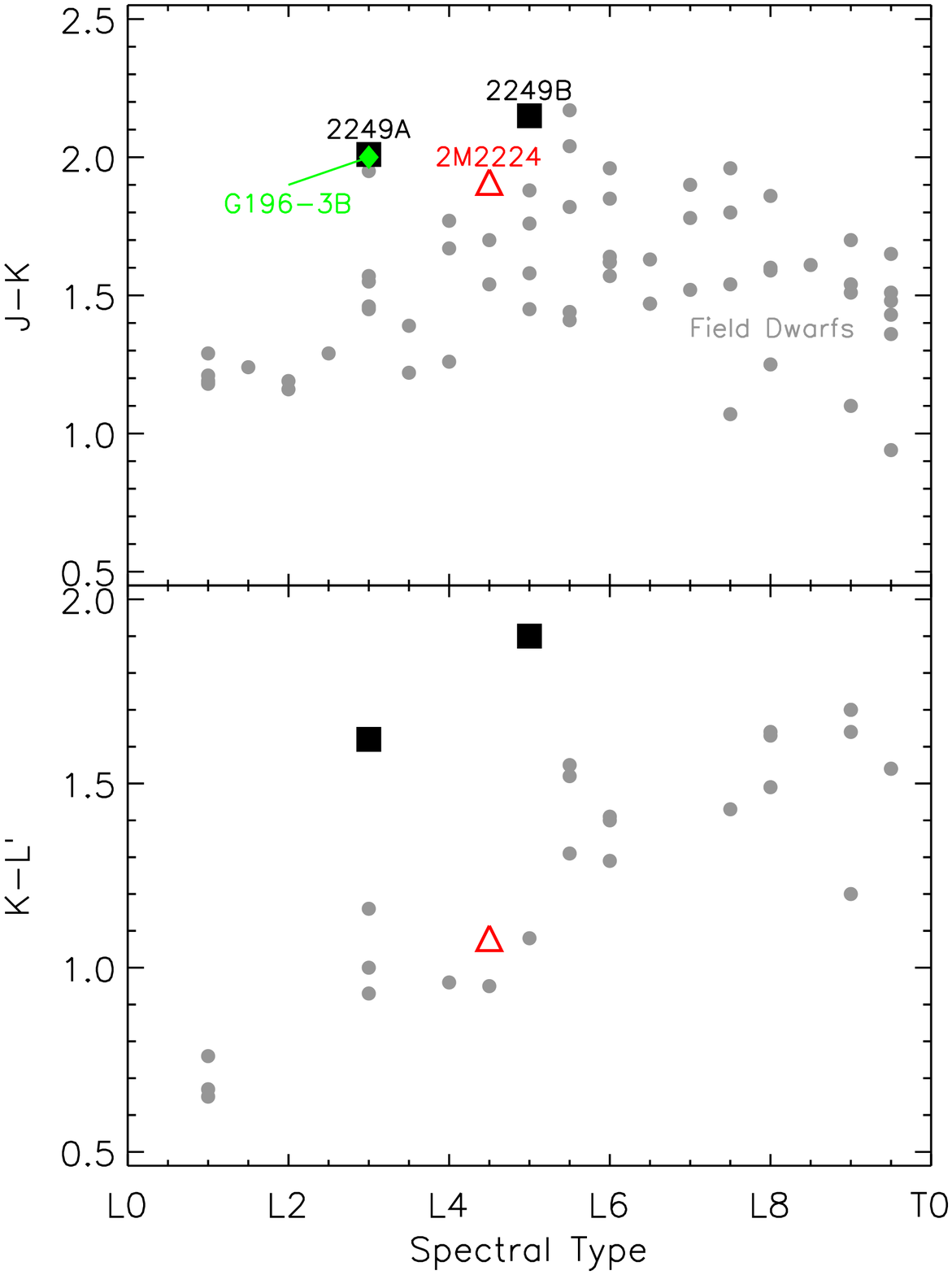}}
\vskip 2ex
\caption[NIR colors]
{\label{colors} MKO $J-K$ (top) and $K-$\Lp\ (bottom) colors as a function of spectral type for L dwarfs.  Field dwarf $J-K$ colors \citep{knapp04} and $K-$\Lp\ colors \citep{golimowski04} are shown as gray circles. The black squares show the colors of \sciencebina\ and B (Table \ref{binphot}).  The green diamond shows the color of G196--3B, a $\sim$100~Myr L3 dwarf \citep{cruz09}, synthesized from its near-IR spectrum.  The open red triangles show the colors of 2MASS~~2224$-$0158 \citep{knapp04,golimowski04}, an L4.5 field dwarf with a dusty photosphere \citep{kirkpatrick00, cushing05}. 
\sciencebina\ and B have red colors compared to normal field dwarfs of the same spectral type, which could be due to a low surface gravity (youth) or a dusty photosphere.}
\end{figure}

As noted by \citet{leggett02} and \citet{knapp04}, \sciencebin\ has unusually red colors ($J-K$=2.05 mag) compared to the other L5 dwarfs in their samples ($J-K$ of 1.41--1.88 mag). 
%, even for the L5 composite spectral they assume.  
Resolved as a binary with spectral types of L3 and L5, the red near-IR colors of \sciencebina\ and B become even more dramatic relative to field dwarfs of the same spectral type (Figure \ref{colors}).
There are currently two explanations for unusually red $J-K$ colors of single L dwarfs: youth \citep{kirkpatrick06} and/or a metal-rich (dusty) photosphere \citep{looper08, cushing05}.
Interestingly, though the $J-K$ colors of young and dusty L dwarfs are similar, the $K-$\Lp\ colors of \sciencebina\ and B are significantly redder than the $K-$\Lp\ of the dusty L dwarf 2MASS~2224-0158 \citep{golimowski04}.  This hints that young and dusty objects may be distinguished on the basis of their $K-$\Lp\ colors, though additional \Lp\ measurements are necessary to establish a set of robust color selection criteria.

Figure \ref{compnir} shows the composite near-IR spectrum of \sciencebin\ compared to an L4.5 dwarf with a dusty photosphere \citep[2MASS~2224-0158;][]{cushing05}, a low-gravity (young) L4 field dwarf \citep[2MASS~0501-0010;][]{cruz09}, a young L3 field dwarf \citep[G196-3B;][]{rebolo98,kirkpatrick01,cruz09}, and a ``normal'' field L3 \citep[2MASS~1506+1321;][]{burgasser07a}.  
The spectrum of \sciencebin\ shows several hallmarks of low gravity.  The FeH bands (particularly at 1.0 and 1.55 $\mu$m) in the spectrum of \sciencebin\ are weaker than seen in the dusty or normal field dwarfs.  
The VO band (1.05~$\mu$m) is stronger in \sciencebin\ than in the older field dwarfs.  
Though the \ion{Na}{1} and \ion{K}{1} features are blended with other features at this spectral resolution, they appear to be weaker in the spectrum of \sciencebin\ than in the older field objects.  
\citet{nakajima04} found that (at $R\simeq$400) the \ion{K}{1} lines are weaker in their $J$-band spectrum of \sciencebin\ than in field dwarfs of similar spectral type.  
The spectrum of \sciencebin\ also shows the triangular $H$-band continuum shape characteristic of young objects \citep[e.g.,][]{lucas01, kirkpatrick06}.  

Though the spectrum of \sciencebin\ shows hallmarks of low-gravity (youth), it is not possible to obtain a precise age for the system.  
\sciencebin\ does not have a measured parallax, so using evolutionary models to determine its age is not possible.  
There are also no known age fiducials for mid-L-type objects showing spectral signatures of youth.  
A strict upper limit on the age of \sciencebin\ can be determined from the youngest L dwarf showing \emph{no} spectral signatures of youth.  HD~130948BC, an L4+L4 binary, has a well established age of 790$^{+220}_{-150}$~Myr \citep{dupuy09}, and its near-IR spectra show no hallmarks of low-gravity \citep{goto02, potter02}.\footnote{We note that our upper age limit is significantly older than that of \citet{kirkpatrick08}, who suggest that spectral signatures of youth in low-resolution optical spectra should be readily discernable for objects Pleiades age and younger ($\lesssim$100~Myr).  We note, however, that \citet{kirkpatrick08} detect subtle signatures of youth in the optical spectrum of Gl417BC, an L4.5+L4.5 dwarf companion system to Gl417A (a G0 dwarf). Using the method described in detail by \citet{dupuy09}, we calculate an age for Gl417A of 750$^{+140}_{-120}$~Myr from gyrochronology \citep{barnes07,mamajek08}.
%, whereas \citet{kirkpatrick01} estimate an age of 80--300~Myr from model isochrones.  
Thus, the $\sim$100~Myr upper age limit of \citet{kirkpatrick08} is possibly an underestimate, and L-type objects displaying spectral signatures of youth in the optical could be significantly older.  We note that the near-IR spectrum of Gl417BC \citep{reid01a,testi01} does not show the same strong IR signatures of youth as \sciencebin, suggesting that \sciencebin\ is likely younger.}
%, whereas Ursa Major ($\sim$400 Myr) L dwarfs show no spectral evidence of low gravity.  
%We opt to use the more conservative 400~Myr age as the upper age limit for \sciencebin.  
Because \sciencebin\ is not spatially coincident with other known young objects, it has to be old enough that its natal cluster has dispersed.  
For a lower limit on the age, this means \sciencebin\ must be at least as old as the youngest known moving groups, the TW~Hydra and $\beta$~Pic associations \citep[$\sim$12~Myr;][]{mentuch08,zuckerman01}.
%This upper age limit may not be applicable for near-IR low-gravity spectral features, but it does place a rough upper limit of $\sim$100~Myr on the age of \sciencebin.

We can also compare \sciencebin\ to G196-3B, an L3 companion to an M2.5 dwarf with an age of $\sim$100~Myr, but with possible ages spanning 20--300~Myr \citep{rebolo98, kirkpatrick01,shkolnik09}.  
Given the remarkable spectral similarity of \sciencebin\ and G196-3B, we adopt an age of $\sim$100 Myr for \sciencebin, but acknowledge that this age is highly uncertain and could range from 20--300~Myr (the age range of G196-3B).  Ages as young as 12~Myr and as old as 790~Myr, though unlikely, can not be completely ruled out and are the strict upper and lower limits on the age of \sciencebin.  The large uncertainties associated with determining the age of \sciencebin\ highlight the need for observations of benchmark low-gravity L dwarfs with well determined ages.
%Overall, the spectrum of \sciencebin\ is quite similar to that of 2MASS~0501-0010, a young L4.  The strengths gravity sensitive lines indicate that it is likely not quite as young as 2MASS~2208+2921.

\subsection{Bolometric Magnitudes}
To calculate the bolometric flux of \sciencebin, we combined our flux calibrated SpeX spectrum with available photometry (Table \ref{binphot}) to create a 0.75--7.87~$\mu$m spectral energy distribution (SED).  
%We extended the SED to 7.87~$\mu$m by assuming that \sciencebin\ has IRAC colors equal to the mean of five young L3--L4 type field dwarfs \citep{luhman09} and stitching the IRAC colors to the \Lp\ photometry of \sciencebin\ using \Lp--IRAC1 (for field L3--L5 dwarfs) from \citet{leggett07}.  
%To estimate the uncertainty of our estimated IRAC fluxes, we combined the measurement error of the \Lp\ photometry and the standard deviation of young field dwarf IRAC colors.
%\footnote{Given the unusually red $J-K$ color of \sciencebin, it is possible that its mid-IR colors might also be red relative to field dwarfs, in which case our use of field dwarf IRAC colors to extrapolate the SED to 7.87 $\mu$m could lead to an underestimate of the bolometric flux.  To test this, we extended the near-IR SED of 2MASS~2224-0158 (a dusty L dwarf with $J-K$=1.91 mag) using the same method we used to construct the mid-IR SED of \sciencebin.  We then compared the resulting SED to the measured IRAC fluxes of 2MASS~2224-0158, and found that they agreed well, to within our uncertainties.}
For wavelengths greater than 7.87~$\mu$m, we assume a Rayleigh-Jeans spectrum.    
We also extended our SED to shorter wavelengths assuming a linear SED from the flux at $i$ band to zero flux at zero wavelength.
Using a Monte Carlo approach to account for uncertainties, we integrate the SED, and calculate an apparent bolometric magnitude, \mbol=17.71 $\pm$ 0.03 mag for the system.\footnote{\mbol= -2.5 log $f_{\rm{bol}}$ - 18.988 \citep{cushing05} with $f_{\rm{bol}}$ expressed in units of $Wm^{-2}$}  
%Our calculation of m$_{bol}$ does not account for the $\sim$10\% variability of \sciencebin\ \citep{knapp04}. 
%Our measured \mbol\ is in good agreement with \mbol=17.76 $\pm$ 0.13~mag determined using the bolometric correction $BC_K$ of \citet{golimowski04}.  We adopt our calculated \mbol\ of 17.66 $\pm$ 0.03 %mag for \sciencebin.
%BC$_K$ does not change significantly for mid-L dwarfs, thus, the uncertainty is dominated by the 0.13 mag scatter of in the BC$_K$--T$_{eff}$ relation, and not uncertainties in the spectral type of \sciencebin.

To determine the \mbol\ of the individual components, we can use the difference in $BC_K$ values \citep{golimowski04} of the two components and their $K$-band flux ratio.  
For \sciencebina, $BC_K$ varies from 3.32 to 3.34 mag for its L3$\pm$0.5 spectral type.  
\sciencebinb\ (L5$\pm$1) has possible $BC_K$ values of 3.32--3.35 mag.  
Since the two components have essentially the same $BC_K$, the $K$-band flux ratio provides the bolometric flux ratio directly.  
We converted our measured \Ks-band flux ratio (0.883 $\pm$ 0.007 mag) to a $K$-band flux ratio by comparing synthetic photometry (from our spatially resolved OSIRIS spectra) and derive a $K$-band flux ratio of 0.893 $\pm$ 0.007 mag. 
The uncertainties for our synthetic flux ratios were small ($<$0.2\%); thus the combined uncertainty is dominated by uncertainties in our measured \Ks\ flux ratio.  
From the integrated-light \mbol\ (17.71 $\pm$ 0.03 mag) and the $K$-band flux ratio, we calculate \mbol's of 18.11 $\pm$ 0.03 and 19.00 $\pm$ 0.03 mag for \sciencebina\ and B respectively.\footnote{We also computed $\Delta$\mbol\ from the $J$-band flux ratio using an SpT-$BC_J$ relation derived from bolometric luminosities in \citet{golimowski04} and $J$-band magnitudes from \citet{knapp04}.  Unlike $BC_K$, $BC_J$ is strongly dependent on spectral type for mid-L-type dwarfs.  From the range of possible spectral types for \sciencebina\ and B, $\Delta$\mbol\ varies from 0.69 to 0.96 mag, which is consistent with the $\Delta$\mbol\  we determine from the $K$-band flux ratio.}  Though we cannot convert \mbol\ to luminosity without a distance, the difference in log(L$_{\rm{bol}}/L_\odot$), 0.36$\pm$0.02~dex, of the two components is distance independent.

\subsection{Distance}

Since \sciencebin\ shows signatures of youth, it is difficult to determine a spectroscopic distance, as absolute magnitude--SpT relations are derived only for field ($\gtrsim$1~Gyr old) dwarfs \citep[e.g.,][]{liu06,cruz03}.  
%At a given spectral type young objects are brighter than their older field counterparts (ref), thus, a spectroscopic distance determined from a field dwarf relation can provide a lower limit on the distance to \sciencebin.
With an estimated age of $\sim$100~Myr, \sciencebin\ is significantly younger than the field population and may not follow the same dependence of absolute magnitude with spectral type.  Nonetheless, we can use the field dwarf SpT--$BC_K$ and SpT--$M_K$ relations \citep{golimowski04, knapp04} to compare to \mbol's of \sciencebina\ and B and obtain a rough distance estimate.  The expected $M_{\rm{bol}}$'s of L3 and L5 field dwarfs are 14.5$\pm$0.3 and 15.4$\pm$0.3 mag, corresponding to distances of 53$\pm$7~pc for both \sciencebina\ and B.  

%Evolutionary models can provide another distance estimate.  The SpTs of \sciencebina\ and B correspond to effective temperatures (T$_{eff}$) of 1949 and 1698~K using the SpT-T$_{eff}$ relation of \citet{golimowski04}.  For an age of 100~Myr, evolutionary models predict luminosities of XXX and XXX, which would place \sciencebin\ at a distance of $\sim$XXX. Since \sciencebina\ and B are young, it is possible that the field dwarf SpT-T$_{eff}$ relation will underestimate their temperatures (based on results for earlier spectral types \citep{luhman03}), which would cause an overestimation of the distance.

To obtain another estimate of distance, we can compare \sciencebin\ to young objects of known distance.  \citet{jameson08} determined the relationship between $J-K$ color and $M_K$ for several young associations, including the Pleiades.  
From their $M_K$--($J-K$) relation for the Pleiades,\footnote{$M_K$=10.70$\pm$0.04 + 1.81$\pm$0.19 ($J-K-1.5$). } we calculate $M_K$ = 11.6 $\pm$ 0.2 and 11.9 $\pm$ 0.2 mag for \sciencebina\ and B, corresponding to distances of 44 $\pm$ 3 and 58 $\pm$ 5 pc, respectively.  We note that the reddest object used to produce the Pleiades relation of \citet{jameson08} has $J-K$=1.92.  Thus, using their relation for \sciencebina\ and B ($J-K$ of 2.01 and 2.14 mag, respectively) is an extrapolation.
\citet{jameson08} also contain $M_K$--($J-K$) relations for Upper Scorpius ($\sim$5~Myr old) and the Hyades ($\sim$625~Myr old), which provide us with conservative upper (71 $\pm$ 7 and 93 $\pm$ 11~pc for \sciencebina\ and B respectively) and lower limits (25 $\pm$ 2 and 33 $\pm$ 3~pc) on the distance.

\begin{figure}
\vskip 0.5in
\centerline{\includegraphics[width=4in]{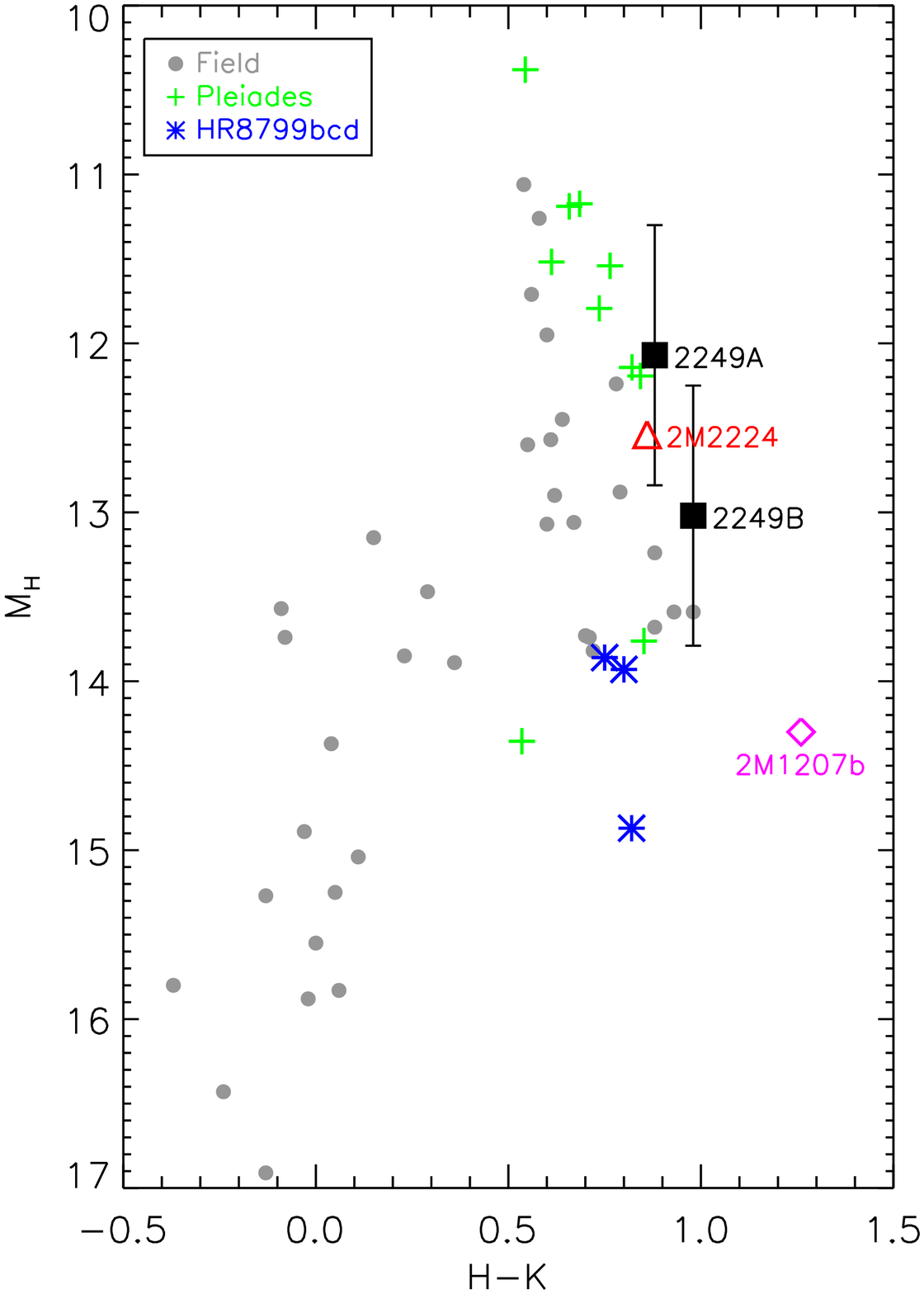}}
\vskip 2ex
\caption[Color Magnitude Diagram]
%{\label{cmd} MKO $H-K$ color versus H-band absolute magnitude.  The black squares show photometry of \sciencebina\ and B assuming a distance of 54$\pm$16~pc.  Field objects \citep{knapp04} are displayed as gray circles and Pleiades candidates from \citet{casewell07} are displayed as green crosses.  For reference, we also plot photometry of the HR8799 planets\citep[blue asterisks; ][]{marois08}, the young planetary-mass object 2MASS~1207b \citep[][]{chauvin04, mohanty07,biller07}, and the dusty L dwarf 2MASS~2224-0158 \citep[red triangle; ][]{knapp04}.  \sciencebin\ appears to follow the Pleiades sequence, which supports our age and distance determination.  We note that the field dwarfs in our plot have M$_H$ values that are $\sim$0.5-1.5 magnitudes fainter than those displayed in a similar plot by \citet{marois08}.  This is due to an error in the generation of their plot (Barman,{\it priv. comm}).  }
{\label{cmd} MKO $H-K$ color versus $H$-band absolute magnitude for L and T~dwarfs. The
black squares show photometry of \sciencebina\ and~B based on our
photometric distance estimate of 54$\pm$16~pc. Field objects
\citep{knapp04} are displayed as gray circles and Pleiades candidates
from \citet{casewell07} are displayed as green crosses. For reference,
we also plot photometry of the HR~8799 planets \citep[blue asterisks;
][]{marois08}, the young planetary-mass object 2MASS~1207b
\citep[magenta diamond;][]{chauvin04, mohanty07,biller07}, and the dusty field L~dwarf
2MASS~2224$-$0158 \citep[red triangle;][]{knapp04}. \sciencebin\ appears to follow 
the Pleiades sequence, which supports our age
and distance determination. Note that the young
planetary-mass objects are somewhat underluminous compared to the field
sequence, though not as dramatically as indicated by a similar version
of this plot in \citet{marois08} (the difference in the two plots
arises from a plotting error in the field object absolute magnitudes in the Marois
\etal\ plot (T. S. Barman, private communication).).}
\end{figure}

Another young object to which we can compare is G196-3B, which has a distance estimate of 21 $\pm$ 6~pc based on the spectroscopic distance to G196-3A \citep{rebolo98}.
%The spectral type (L3) and $J-K$ color of G196-3B (2.00 $\pm$ 0.06 mag) are nearly identical to those of \sciencebina\ (L3$\pm$0.5, 2.01 $\pm$ 0.08 mag)
%, thus we can compare their $K$ magnitudes to obtain a distance estimate.
%G196-3B has an absolute $K$ magnitude, $M_K$=11.1 $\pm$ 0.7 mag.\footnote{To compare our (MKO) photometry of \sciencebin\ to G196-3B, we first synthesized MKO $JHK$ photometry from our spectrum of G196-3B normalized to its 2MASS~ASS $JHK$ photometry.  
%We derive MKO $JHK$ magnitudes of 14.73$\pm$0.05, 13.73$\pm$0.05, and 12.73$\pm$0.04 mag for G196-3B.}  
%In order for \sciencebina\ to have the same value of M$_K$ it would have to be at a distance of 56$\pm$19~pc.  
The spectral type (L3) and near-IR colors of G196-3B ($J-K$=2.00 $\pm$ 0.06 mag) are nearly identical to those of \sciencebina\ (L3$\pm$0.5, 2.01 $\pm$ 0.08 mag) thus we can compare their near-IR magnitudes to calculate their relative distances.\footnote{To compare our (MKO) photometry of \sciencebin\ to G196-3B, we first synthesized MKO $JHK$ photometry from our spectrum of G196-3B normalized to its Two Micron All Sky Survey (2MASS) $JHK$ photometry.  We derive MKO $JHK$ magnitudes of 14.73$\pm$0.05, 13.73$\pm$0.05, and 12.73$\pm$0.04 mag for G196-3B.}  Assuming the two objects have the same age, \sciencebina\ is 2.59$\pm$0.05 times more distant than G196-3B.  Scaling from the distance estimate for G196-3B, we estimate that \sciencebina\ has a distance of 54$\pm$16~pc.
Our distance estimates from comparison to field dwarfs, Pleiades objects, and G196-3B agree to within their uncertainties.  
We adopt the most conservative of our distance estimates (54 $\pm$ 16~pc) as the distance to \sciencebin, with the caveat that this distance could be an underestimate if the system is significantly younger than the Pleiades or G196-3B.   Figure \ref{cmd} shows that \sciencebina\ and B roughly follow the Pleiades sequence, which supports our distance determination.

At a distance of 54$\pm$16~pc, the projected physical separation of \sciencebin\ is 17$\pm$5~AU, and the luminosities (log($L_{\rm{bol}}/L_\odot$)) of \sciencebina\ and B are $-3.9 \pm 0.3$ and $-4.2 \pm 0.3$~dex.

\subsection{Mass}

\begin{figure}
\vskip 0.5in
\centerline{\includegraphics[width=5in]{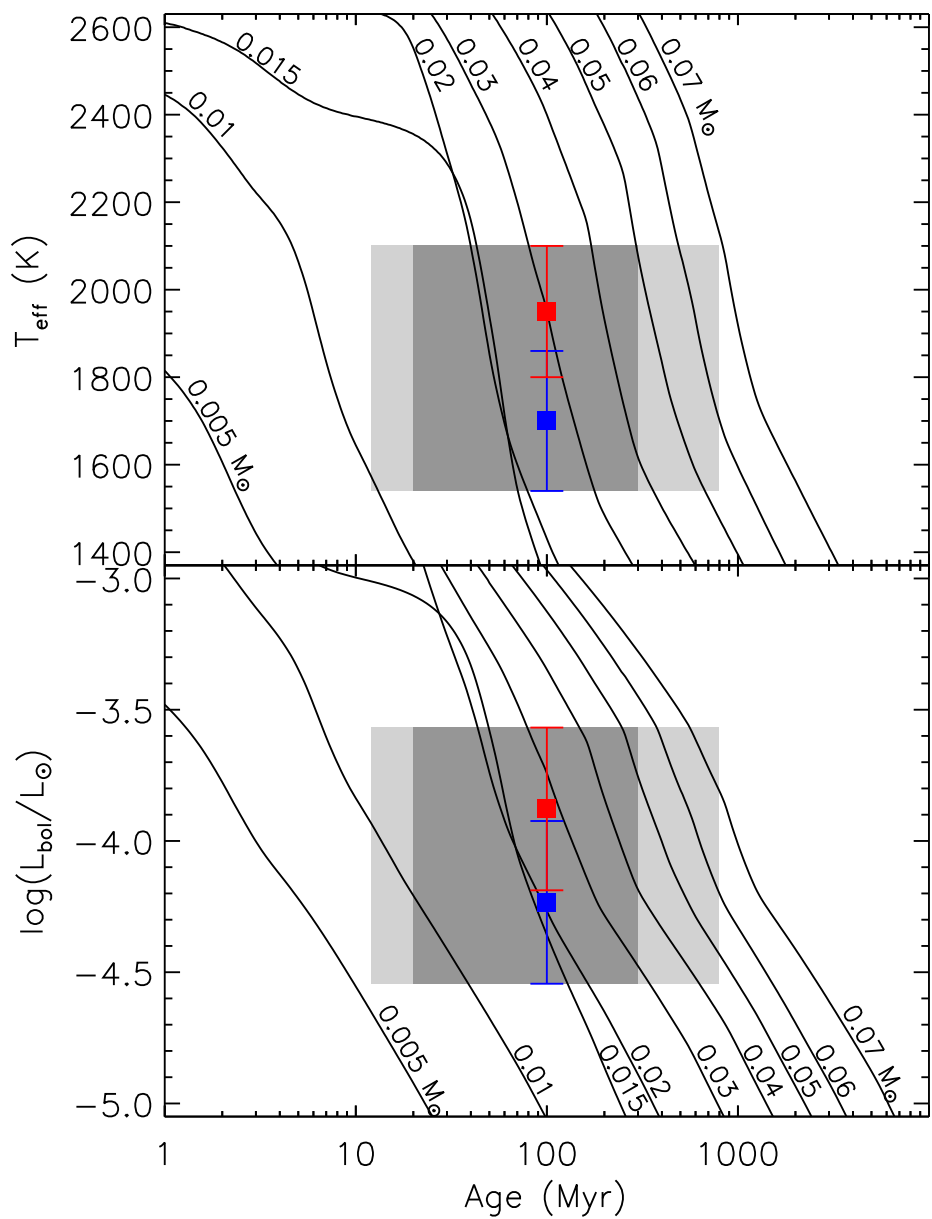}}
\vskip 2ex
\caption[Mass determination]
{\label{hr} Plot of age vs. effective temperature (top) and log($L_{\rm{bol}}/L_\odot$) (bottom) for \sciencebina\ and B with model iso-mass contours from \citet{burrows97} overlaid.  The shaded regions show the range of possible ages for \sciencebin\ from our strict age limits (12-790~Myr; light gray shading) and the age of G196--3B (20--300~Myr; gray shading).}
\end{figure}

To estimate the mass of \sciencebina\ and B, we compare their luminosities and effective temperatures to predictions of evolutionary models.  
The SpT's of \sciencebina\ and B (L3.0$\pm$0.5 and L5$\pm$1) correspond to effective temperatures ($T_{eff}$) of 1950$\pm$140 and 1700$\pm$160~K using the $SpT-T_{eff}$ relation of \citet{golimowski04},\footnote{Since \sciencebina\ and B are young, it is possible that the field dwarf SpT--$T_{\rm{eff}}$ relation will underestimate their temperatures (based on results for young objects with earlier spectral types \citep{luhman03}).} with uncertainties derived from the scatter of the relation (124~K) and the uncertainty in SpTs.  
Figure \ref{hr} shows age versus $T_{eff}$ and log($L_{\rm{bol}}/L_\odot$) for \sciencebina\ and B with the evolutionary models of \citet{burrows97} overlaid. 
At an age of 100~Myr, the evolutionary models predict masses of 0.029$\pm$0.006 and 0.022$^{+0.006}_{-0.009}$~$M_\odot$ from the luminosities of \sciencebina\ and B respectively.\footnote{Uncertainties in the mass are determined from the uncertainties in luminosity (or $T_{\rm{eff}}$), and do not include age uncertainties or possible systematic errors in the evolutionary models \citep{dupuy09}.}  From the $T_{\rm{eff}}$'s of \sciencebina\ and B, the evolutionary models predict masses of 0.030$^{+0.003}_{-0.002}$ and 0.026$^{+0.002}_{-0.003} M_\odot$. The masses predicted from $T_{\rm{eff}}$ and luminosity are in good agreement, which is reassuring given that the luminosities were determined from the photometry of the system, whereas the $T_{eff}$'s are determined from the SpTs of \sciencebina\ and B.  Given the estimated masses of \sciencebin, the system will likely evolve into a late-T/Y dwarf system \citep{liu10} when it reaches the typical age of the field population ($\sim$1~Gyr).
%Our mass estimates do not include uncertainty i that the evolutionary models can have significant systematic errors \citep{dupuy09}.

The uncertain age of \sciencebin\ adds additional uncertainty to the mass determinations. 
For an age range of 20--300~Myr, the evolutionary models predict masses ranging from 0.011 to 0.056~$M_\odot$ and from 0.008 to 0.050~$M_\odot$ for \sciencebina\ and B based on their luminosities.  Over the same age range, their $T_{\rm{eff}}$'s yield possible masses of 0.013--0.050~$M_\odot$ and 0.012--0.045~$M_\odot$.  For our strict age limits (12--790~Myr), the mass of \sciencebina\ could range from 0.011 to 0.070~$M_\odot$ and the mass of \sciencebinb\ could range from 0.009--0.065~$M_\odot$.
Thus, although their masses are very uncertain, both components are likely substellar ($<$0.072~$M_\odot$).

\begin{figure}
\vskip 0.5in
\centerline{\includegraphics[width=6in,angle=90]{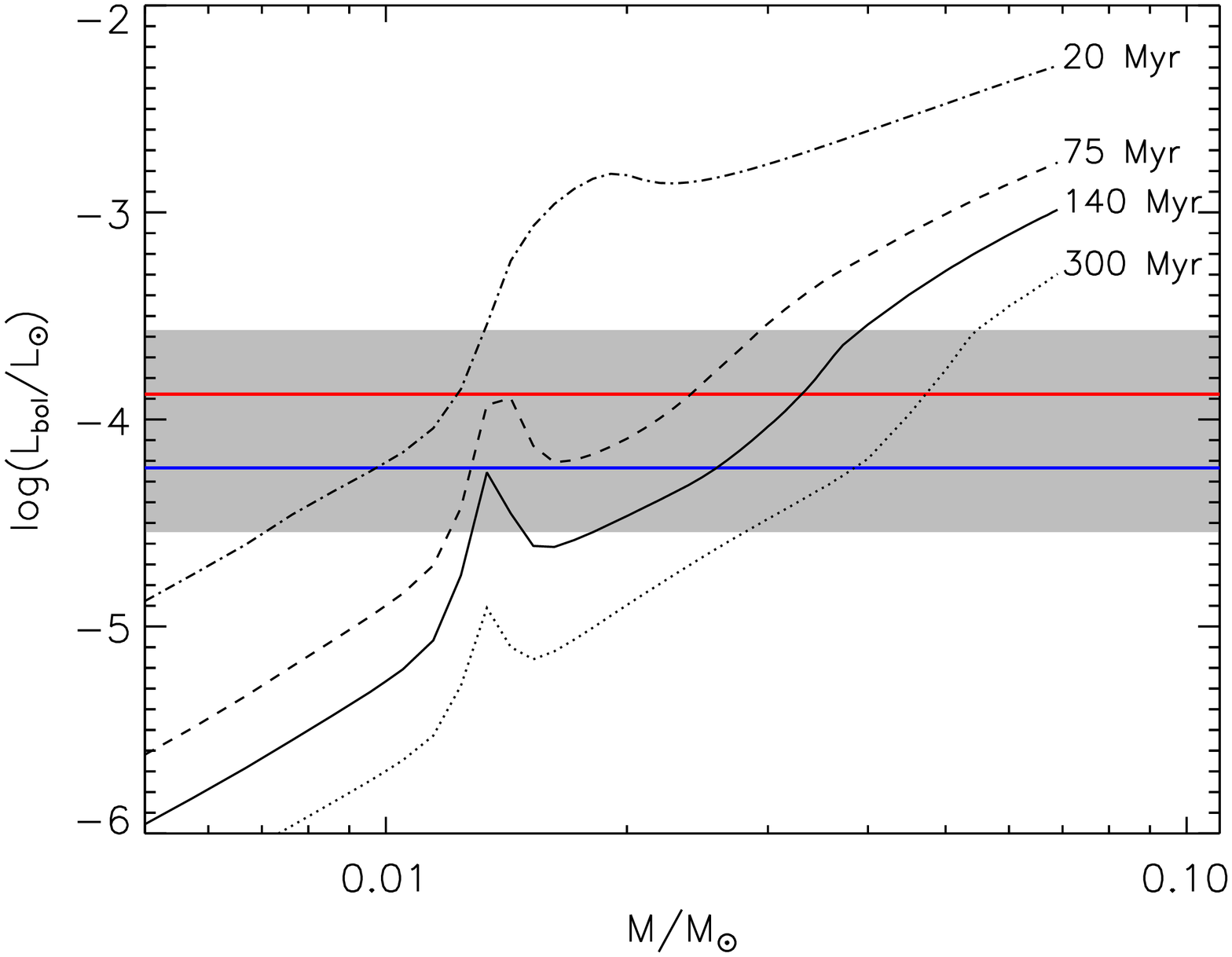}}
\vskip 2ex
\caption[Mass Ratio determination]
{\label{massratio} Plot of mass vs. log($L_{\rm{bol}}/L_\odot$) for model isochrones of 20, 75, 140, and 300~Myr \citep{burrows97}.  The solid horizontal lines show our measured luminosities of \sciencebina\ (red) and B (blue), while the shaded region represents the range of 1$\sigma$ uncertainty.  While the absolute luminosities of \sciencebina\ and B are quite uncertain, the relative luminosity (represented by the separation between the red and blue lines; 0.36 $\pm$ 0.02~dex) is well determined. The bump in the isochrones at $\sim$0.013~$M_{\odot}$ is due to deuterium burning and means that for the range of possible ages and luminosities of \sciencebin, its mass ratio cannot be robustly determined.  In fact, \sciencebin\ could have a flux reversal due to deuterium burning, where the lower mass object is currently burning deuterium and is thus the more luminous component.}
\end{figure}

$\Delta$log($L_{\rm{bol}}/L_\odot$) can be used to infer the mass ratio of a system using the analytic scaling relation from \citet{burrows01}:  $L_{\rm{bol}} \propto M^{2.64}t^{-1.3} \kappa_R^{0.35}$.
Using this relation, the well-determined $\Delta$log($L_{\rm{bol}}/L_\odot$)=0.36 $\pm$ 0.02 of \sciencebin\ corresponds to a mass ratio, $q \equiv M_B/M_A = 0.74 \pm 0.01$, which is lower than any known L dwarf binary.\footnote{http://www.vlmbinaries.org}  It is important to note, however, that the \citet{burrows01} power-law relations are meant to describe the late-time evolution of brown dwarfs and do not fully describe younger objects.  Figure \ref{massratio} shows isochrones from the evolutionary models of \citet{burrows97}.  The bump in the isochrones at masses of $\sim$0.01--0.02 $M_{\odot}$ is due to deuterium burning and means that $\Delta$log($L_{\rm{bol}}/L_\odot$) can not be used to accurately infer the mass ratio if either component is currently burning deuterium.  For instance, if \sciencebinb\ (the less luminous component) is currently burning deuterium (age of ~$\sim$140~Myr), the mass ratio of the system could be as low as 0.4.  On the other hand, if \sciencebina\ (the more luminous component) is burning deuterium (age of $\sim$75~Myr), the mass ratio would be closer to unity, and the system could exhibit a reversal in the luminosities of the binary, with \sciencebina\ being the lower mass component.  The effect of deuterium burning on the evolution of brown dwarfs is discussed in detail by \citet{saumon08}. Unfortunately, without better constraints on the age and luminosities of \sciencebina\ and B, we can not distinguish between the various mass ratio possibilities.  \sciencebin\ is the first known binary system that could exhibit a flux reversal induced by deuterium burning.
%We can, however, use the fact that the system is coeval and our well determined value of $\Delta$log(L_\rm{bol}/L$_\odot)$ (0.35$\pm$0.02) to constrain the mass ratio.  
%;Figure \ref{massratio} shows power law fits to log(L/L$_\odot)$ and mass for isochrones of 10, 50, 100, and 500 Myr \citep[DUSTY models,][]{chabrier00}.  
%;The power law index varies from 0.36 to 0.42 corresponding to mass ratios of 0.71--0.75.  
%;The mass ratio of \sciencebin\ is lower than any known L-dwarf binary.

\subsection{Orbital Period}
We have measured the projected separation of \sciencebin\ to be
322.8$\pm$0.3~mas, but the true semimajor axis of the binary depends on its
orbital parameters.  Following the method of
\citet{torres99}, we assume random viewing angles and a
uniform eccentricity distribution between 0~$<~e~<$~1 to derive a
correction factor of 1.10$^{+0.90}_{-0.36}$ (68.3\% confidence limits)
to convert projected separation into a semimajor axis.  At a distance of
54$\pm$16~pc, this results in a semimajor axis of
19$^{+17}_{-8}$~AU.  For a total mass of 0.051~$M_\odot$ (assuming an age of $\sim$100~Myr old), this corresponds
to an orbital period of 400$^{+600}_{-200}$~years. 
%For a total mass of 0.095~\Msun (the total mass of \sciencebin\ if 300~Myr old), this corresponds to an orbital period of 193$^{+235}_{-130}$~years. 
Unfortunately, the estimated orbital period of \sciencebin\ is too long for our two epochs of observations (Table \ref{tbl:obs}) to detect any relative motion of the binary.  
%Nonetheless, future astrometric monitoring of \sciencebin\ could yield 

%From the masses (for an age of 100~Myr) and semi-major axis of \sciencebin, we calculate a binding energy of $\sim$9$~\times 10^{41}$~erg for \sciencebin, which is lower than all but two known field binaries.  The total mass of \sciencebin\ is  

\subsection{A Common Proper Motion M Dwarf}
\indent \sciencebin\ lies within the SDSS region known as Stripe 82,
which is a 2$\fdg$5 wide strip spanning 99\degs\ of the celestial
equator.  \citet{bramich08} analyzed all available observations of
this region spanning the seven years of survey operations from 1998 to
2005, computing light curves and proper motions on an extragalactic
reference frame from catalog data.  They computed a
proper motion of  $\mu = 82.8\pm3.7$~mas~yr$^{-1}$ at a P.A. of
$85\fdg4\pm2\fdg5$ for \sciencebin.\footnote{We note that the proper motion calculated by \citet{bramich08} agrees to within the uncertainties with other recent measurements \citep{scholz09,faherty08}.}  We searched the light-motion catalog for
any other objects with a well-detected proper motion
($\sigma_{\mu}/\mu > 10$ and $\sigma_{\rm PA} < 10\deg$) that share
similar motion (both $\mu$ and PA within 3$\sigma$) with \sciencebin.
%(i.e., both $\mu$ and PA within 3$\sigma$
%of the measured values for \sciencebin).  
We found one such object
within a limit of 700\arcsec\ of \sciencebin\ ($3.8\times10^4$~AU at 54~pc,
i.e., well beyond the wide binary cutoff of $10^4$~AU found in
simulations by \citet{weinberg87}).  
This common proper motion
companion is GSC~00568-01752, which lies $48\farcs937\pm0\farcs009$ at a PA of $257\fdg04\pm0\fdg01$ away from \sciencebin\ (a projected separation of 2600 $\pm$ 800~AU for d=54 $\pm$ 16~pc) and has a proper motion of  $\mu$ =$90.7\pm3.3$~mas~yr$^{-1}$ and PA=$87\fdg0\pm2\fdg1$, in agreement with the proper motion of \sciencebin\ at
1.6$\sigma$ and 0.5$\sigma$, respectively.

By comparing the SDSS and 2MASS photometry of GSC~00568-01752\footnote{$r$=13.63$\pm$0.01, $i$=12.45$\pm$0.01, $z$=11.93$\pm$0.01, $J$=10.35$\pm$0.03, $H$=8.74$\pm$0.03, and $K$=9.51$\pm$0.023 mag.} with colors for field M dwarfs \citep{west08}, we determine a spectral type of M2$\pm$2 via $\chi^2$ minimization.  From the $i-z$ photometric parallax relation of \citet{west05}, we calculate a distance of 53$\pm$15~pc for GSC~00568-01752, in excellent agreement with our distance estimate for \sciencebin.  GSC~00568-01752 has a low S/N X-ray detection in the \emph{ROSAT} All-Sky Survey \citep[1RXS~J224957.2+004402;][]{mickaelian06} from which we calculate $F_{\rm{X}}$ using the count rate to flux conversion of \citet{schmitt95}.  We determine log($F_{\rm{X}}/F_J$)=-2.3$\pm$0.3\footnote{Our calculation of $F_J$ uses a $J$-band flux density for Vega of 3.129$\pm$0.055 $\times 10^{-13}$~W~cm$^{-2}$~$\mu$m$^{-1}$ \citep{cohen03} and a bandwidth of 0.29 $\mu$m} for GSC~00568-01752 which is consistent with values for young ($\lesssim$650 Myr old) nearby M dwarfs \citep{shkolnik09} and provides additional evidence that GSC~00568-01752 and \sciencebin\ are associated.  Further study of GSC~00568-01752 could help to constrain the age and distance of \sciencebin.

\subsection{Moving Group Membership}
\begin{figure}
\vskip 0.5in
\centerline{\includegraphics[width=4in]{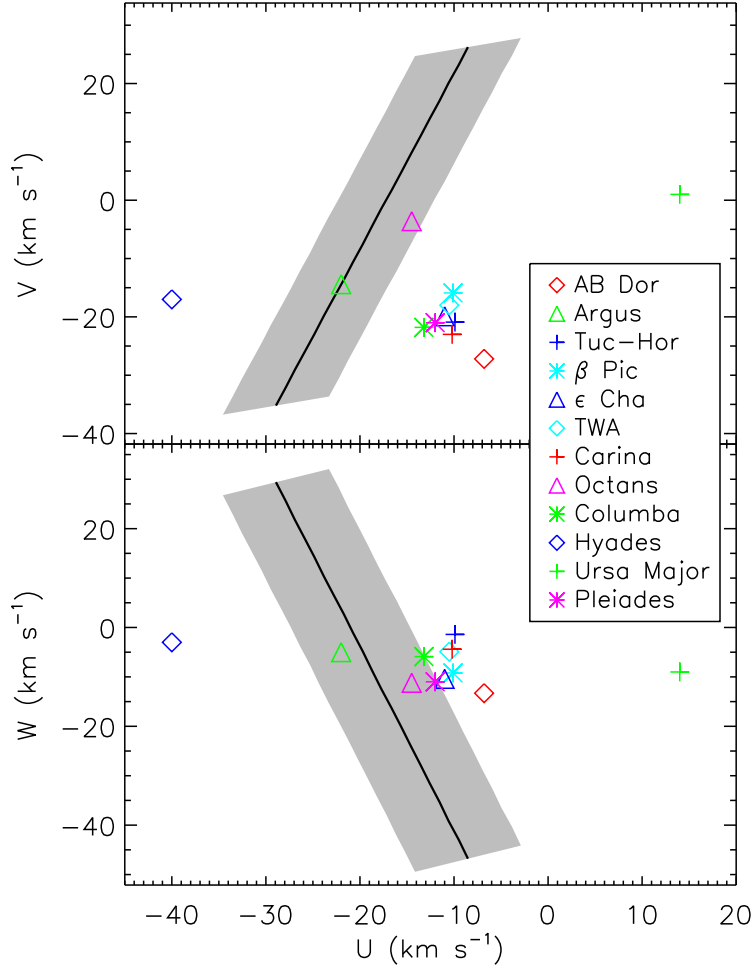}}
\vskip 2ex
\caption[Galactic Space Motion]
{\label{uvw} $UVW$ space for \sciencebin\ and known young moving groups, plotted using a right-handed coordinate system ($U$ positive toward the Galactic center).  $UVW$ for \sciencebin\ (solid black line) is calculated from its position, distance, and proper motion for a range of radial velocities (-50 to 50~km~s$^{-1}$).  The gray shaded region shows the region of 1$\sigma$ uncertainty.  Note that a broader range of radial velocities would extend the length of the solid black line, but the width of the 1$\sigma$ uncertainty would remain unchanged.  The $UVW$s of moving groups come from \citet{torres08} and \citet{montes01}.  The range of possible space motion for \sciencebin\ does not overlap most of the known young moving groups with the exception of the Octans and Argus associations, but these associations are in a different part of the sky compared to \sciencebin\ (Figure \ref{ykg}).}
\end{figure}

\begin{figure}
\vskip 0.5in
\centerline{\includegraphics[width=5in,angle=90]{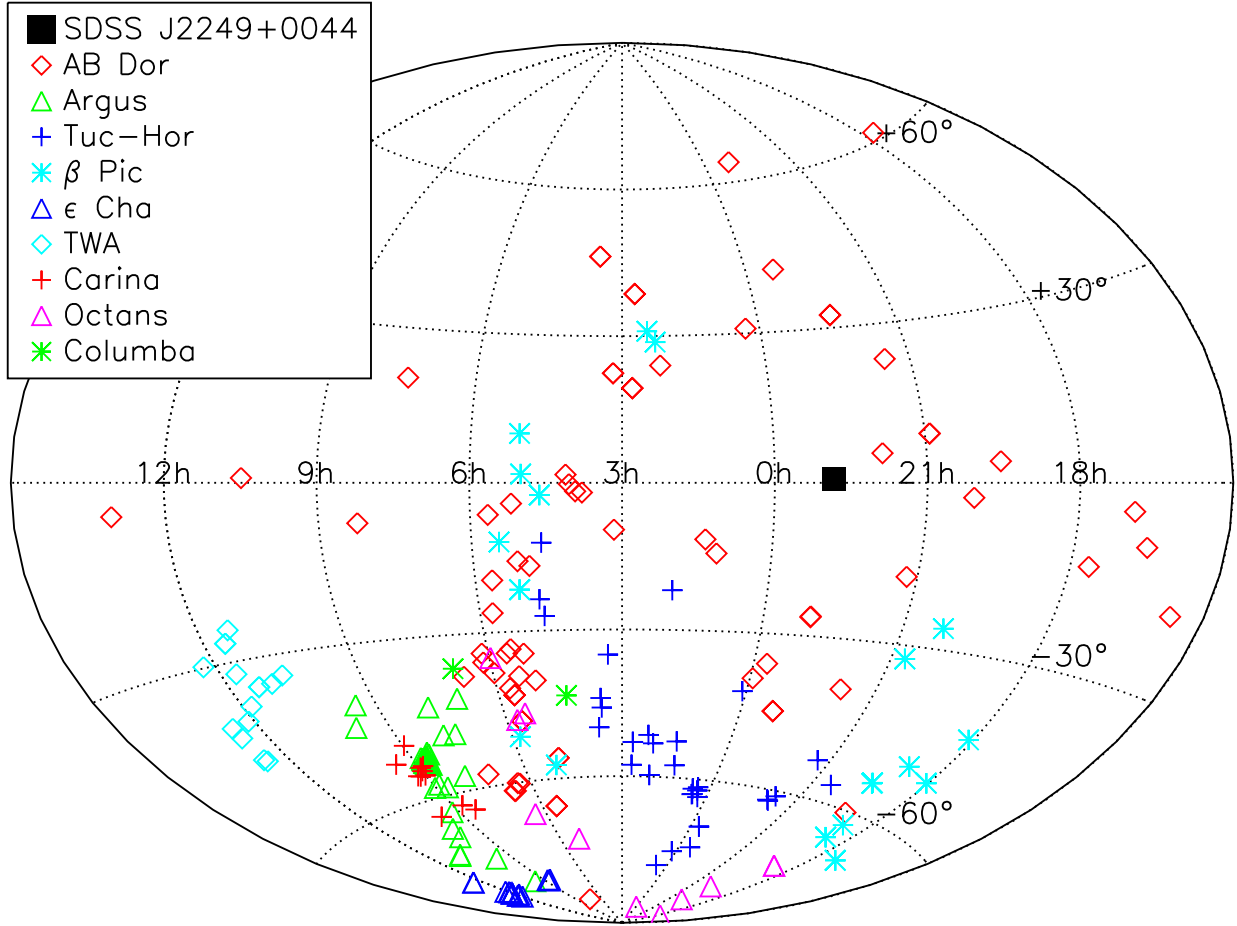}}
\vskip 2ex
\caption[YKG coordinates]
{\label{ykg} Aitoff projection (J2000 equatorial coordinates) showing the positions of \sciencebin\ and young moving group members from \citet{torres08}. Argus and Octans, the two young moving groups having space motions that overlap with the possible space motion of \sciencebin, are concentrated deep in the southern hemisphere.  Membership of \sciencebin\ in one of these associations is thus unlikely.}
\end{figure}

Young moving groups and associations are loose groups of stars sharing common ages, compositions, and kinematics.  
Thus, membership of \sciencebin\ in a young moving group would provide an independent estimate of age and metallicity.
Membership in a young moving group is typically confirmed via common space motion ($UVW$), which requires knowledge of the object's distance, proper motion, and radial velocity.  
Using the photometric distance estimate of \sciencebin\ (54$\pm$16~pc) and its proper motion \citep[$\mu = 82.8\pm3.7$~mas~yr$^{-1}$ and PA=$85\fdg4\pm2\fdg5$;][]{bramich08}, we calculated its range of possible $UVW$ for radial velocities from -50 to 50 km~s$^{-1}$.
%Given the distance to \sciencebin\ and its implied youth, could it be a member of a young moving group? 
Figure \ref{uvw} shows the possible space motion of \sciencebin\ compared to known moving groups \citep{torres08,zuckerman04}.  Only two groups, the Octans ($\sim$20~Myr old) and Argus ($\sim$40~Myr old) associations, have $UVW$ that overlap the 1$\sigma$ range of \sciencebin.  Both of these young groups, however, are concentrated far in the southern hemisphere (Figure \ref{ykg}), making membership with \sciencebin\ unlikely.
%side note: no value for d (from 10-100 pc) can get it within the "good" box (i.e. 10 kms from the local association).
%We note that a number of known moving groups have space motions which fall within the 2$\sigma$ range of \sciencebin.
\sciencebin\ could be an isolated young field object.  Radial velocity and parallax measurements for \sciencebin\ are necessary to further assess possible membership with young moving groups.

\section{Conclusions}
As a part of our Keck LGS~AO survey of young field brown dwarfs, we have discovered the young L dwarf, \sciencebin\, to be a 0$\farcs$32 binary.  
Over our two epochs of imaging spanning nearly two years, the separation and P.A. of the system do not change significantly, indicating that \sciencebina\ and B are comoving and thus physically associated.   
Using spatially resolved $K$-band spectroscopy, we determine spectral types of L3$\pm$0.5 and L5$\pm$1 for \sciencebina\ and B, respectively.  \sciencebina\ and B have unusually red near-IR colors relative to field dwarfs, which can indicate either a young age or a dusty photosphere.  The FeH, \ion{K}{1}, and \ion{Na}{1} absorption features in the near-IR spectrum of \sciencebin\ are weaker and the VO absorption is stronger than seen in normal and dusty photosphere L dwarf spectra, indicating a young age for \sciencebin.  The $K$--\Lp\ colors of \sciencebina\ and B are significantly redder than 2MASS~2224-0158, a dusty L4.5 dwarf, hinting that $K$--\Lp\ color can be used to distinguish between young and dusty objects.
The near-IR spectra of \sciencebin\ and G196-3B (a young L3 dwarf) are remarkably similar, thus we adopt an age of 100~Myr for \sciencebin, but note that ages of 12-790~Myr are possible.  By comparing the $K$-band magnitude of \sciencebina\ to that of G196-3B (which has a distance estimate from the spectroscopic distance of G196-3A), we estimate a distance of 54$\pm$16~pc for the system.

%Using our near-IR spectrum and available optical and infrared photometry, we calculate the bolometric luminosity of \sciencebina\ and B.  Though the uncertain distance to the binary corresponds to large uncertainties in the absolute luminosities, the difference in log(L/L$_\odot)$, 0.36$\pm$0.02~dex, of the two components is distance-independent.  
Comparison of the luminosities of \sciencebina\ and B to evolutionary models for an age of 100~Myr yields masses of 0.029$\pm$0.006 and 0.021$^{+0.006}_{-0.009}$~$M_\odot$ respectively.  For ages of 12--790~Myr, the masses could range from 0.011 to 0.070~$M_\odot$ for \sciencebina\ and from 0.009 to 0.065~$M_\odot$ for \sciencebinb.  Thus, \sciencebina\ and B are clearly substellar.  Evolutionary models predict that for the range of possible ages and luminosities either component could be burning deuterium.  The mass ratios of the binary could range from 0.4 to near unity depending on which component is burning deuterium.  The system is the first known binary which could exhibit a flux reversal due to deuterium burning.  For a range of plausible radial velocities, we show that the position, proper motion, and distance to the system are inconsistent with membership in known young moving groups.  A common proper motion search of SDSS stripe 82 shows that GSC~00568-01752 (which has colors consistent with an early M dwarf) is likely associated with \sciencebin, and further study of this M dwarf could help to better contrain the age, distance and kinematics of \sciencebin.  As a young, low-mass system having very red near-IR colors, \sciencebin\ is a valuable template for future studies of planetary-mass objects.  Better constraints on the age, luminosities, and effective temperatures of \sciencebina\ and B could provide critical tests to evolutionary models for young L dwarfs.

\acknowledgments
We gratefully acknowledge the Keck LGS AO team for their exceptional
efforts in bringing the LGS AO system to fruition.  It is a pleasure
to thank Randy Campbell, Jim Lyke, Al Conrad, Hien Tran, Christine Melcher, Cindy Wilburn, Joel Aycock, Jason McElroy, and the Keck Observatory staff for assistance
with the observations.  This publication benefitted from helpful discussions with Kevin Marshall and Evgenya Shkolnik regarding \emph{ROSAT} data.  We sincerely thank Eric Mamajek for informing us about GSC~00568-01752, the possible M dwarf companion to \sciencebin.  We also thank John Rayner and the IRTF Observatory staff for assistance with our SpeX observations and data reduction.  
This publication has made use of the VLM Binaries Archive maintained by Nick Siegler at  http://www.vlmbinaries.org.
This research has benefitted from the M, L, and T dwarf compendium housed at DwarfArchives.org and maintained by Chris Gelino, Davy Kirkpatrick, and Adam Burgasser as well as from the SpeX Prism Spectral Libraries, maintained by Adam Burgasser at http://www.browndwarfs.org/spexprism.  M.C.L., K.N.A. and T.J.D. acknowledge support for this work from NSF grant AST-0507833 and AST-0407441.  M.C.L. also acknowledges support from an Alfred P. Sloan Research Fellowship.  K.N.A. was partially supported by NASA Origins of Solar Systems grant NNX07AI83G.

\clearpage

%%%%%%%%%%%%%%%%%%%%%%%%%%%%%%%%%%%%%%%%
% TABLE 1 - KECK OBSERVATIONS
%%%%%%%%%%%%%%%%%%%%%%%%%%%%%%%%%%%%%%%%
\begin{deluxetable}{lcccccccc}
\tablecaption{Keck LGS AO Observations}
\tabletypesize{\scriptsize}
%\rotate
%\tabletypesize{\footnotesize}
\tablewidth{0pt}
%\tablecolumns{2}
\tablehead{
  \colhead{Date} &
  \colhead{Instrument} &
  \colhead{Airmass} &
  \colhead{Filter} &
  \colhead{FWHM} &
  \colhead{Strehl Ratio\tablenotemark{a}} &
  \colhead{$\rho$} &
  \colhead{P.A.} &
  \colhead{Flux Ratio} \\
  \colhead{(UT)} &
  \colhead{} &
  \colhead{} &
  \colhead{} &
  \colhead{(mas)} &
  \colhead{} &
  \colhead{(mas)} &
  \colhead{(deg)} &
  \colhead{(mag)}
}
\startdata
 2006 Oct 14 & NIRC2  &1.07 &  $J$   &    61$\pm$7    &  0.048$\pm$0.009 & 322.0$\pm$1.0 & 279.53$\pm$0.08 & 1.024$\pm$0.016  \\
             & NIRC2  &1.07 &  $H$   &    56$\pm$6    &   0.13$\pm$0.03  & 322.8$\pm$0.3 & 279.49$\pm$0.03 & 0.953$\pm$0.009  \\
             & NIRC2  &1.06 &  \Ks\  &    57$\pm$3    &   0.31$\pm$0.06  & 322.8$\pm$0.3 & 279.50$\pm$0.04 & 0.883$\pm$0.007  \\
 2008 Sep  8 & NIRC2  & 1.06 &  \Lp\  & 93\tablenotemark{b} & 0.22\tablenotemark{b} & 320.8$\pm$0.8 &  280.9$\pm$0.3  &  0.60$\pm$0.05  \\
 2008 Sep  9 & OSIRIS & 1.35 & $Zbb$  &     130$\pm$20      &        0.015\tablenotemark{b}        &   309$\pm$7   &  281.5$\pm$0.8  &  1.12$\pm$0.04  \\
             & OSIRIS & 1.43 & $Hn$1  &      95$\pm$12      &        0.047\tablenotemark{b}&   307$\pm$6   &  281.0$\pm$0.5  &  1.02$\pm$0.04  \\

\enddata
\tablenotetext{a}{For NIRC2 images, Strehl ratios were computed using
  the publicly available routine \texttt{NIRC2STREHL}.}

\tablenotetext{b}{These values are measured from the stacked mosaic of
  individual dithers and thus do not have rms errors.}

\label{tbl:obs}
\end{deluxetable}

%%%%%%%%%%%%%%%%%%%%Table 2%%%%%%%%%%%%
\begin{deluxetable}{lccc}
\tablecolumns{4}
\footnotesize
\tablecaption{Photometry of \sciencebin}
\tablewidth{0pt}
\tablehead{
\colhead{Band}                  &
\colhead{$A+B$}             &
\colhead{$A$}             &
\colhead{$B$}             \\
\colhead{}                  &
\colhead{(mag)}             &
\colhead{(mag)}             &
\colhead{(mag)}             
}
\startdata 
$i$   & 21.64$\pm$0.05\tablenotemark{a} & \nodata & \nodata \\
$z$   & 19.48$\pm$0.02\tablenotemark{a} & \nodata & \nodata \\
$Z$   & 18.24$\pm$0.05\tablenotemark{b} & \nodata & \nodata \\
$Zbb$ & 17.33$\pm$0.05\tablenotemark{c} & 17.66$\pm$0.05 & 18.78$\pm$0.06 \\
$J$   & 16.47$\pm$0.07\tablenotemark{d} & 16.83$\pm$0.07 & 17.85$\pm$0.07 \\
$Hn1$ & 15.98$\pm$0.05\tablenotemark{c} & 16.34$\pm$0.05 & 17.36$\pm$0.06 \\
$H$   & 15.36$\pm$0.03\tablenotemark{d} & 15.74$\pm$0.04 & 16.69$\pm$0.04 \\ 
$K$   & 14.42$\pm$0.03\tablenotemark{d} & 14.82$\pm$0.03\tablenotemark{e} & 15.71$\pm$0.03\tablenotemark{e} \\
\Lp\  & 12.71$\pm$0.07\tablenotemark{f} & 13.20$\pm$0.07 & 13.80$\pm$0.07 \\
$[3.6]$ & 13.24$\pm$0.05 & \nodata & \nodata \\
$[4.5]$ & 13.07$\pm$0.05 & \nodata & \nodata \\
$[5.8]$ & 12.73$\pm$0.05 & \nodata & \nodata \\
$[8.0]$ & 12.55$\pm$0.05 & \nodata & \nodata \\
\enddata
\tablenotetext{a}{from \citet{scholz09} in SDSS magnitude and filter systems}
\tablenotetext{b}{from \citet{leggett02}}
\tablenotetext{c}{synthesized from our SpeX spectrum}
\tablenotetext{d}{weighted mean of values reported in Table 9 of \citet{knapp04}}
\tablenotetext{e}{\Ks-band flux ratio (0.883 $\pm$ 0.007 mag) converted to $K$-band flux ratio (0.893 $\pm$ 0.007 mag) by synthesizing $K$-band photometry from our OSIRIS spectra}
\tablenotetext{f}{from \citet{golimowski04}}
\label{binphot}
\end{deluxetable}

%%%%%%%%%%%%%%%%%%%%Index Table%%%%%%%%%%%%%%%%%
\begin{deluxetable}{llcccc}
\tablecolumns{6}
\footnotesize
\tablecaption{$K$-band Spectral Indices}
\tablewidth{0pt}
\tablehead{
\colhead{}                  &
\colhead{}   &
\multicolumn{2}{c}{\sciencebina}             &
\multicolumn{2}{c}{\sciencebinb}             \\
\colhead{Index} &
\colhead{Reference}             &
\colhead{Value\tablenotemark{a}}      &
\colhead{SpT\tablenotemark{b}}             &
\colhead{Value\tablenotemark{a}}             &
\colhead{SpT\tablenotemark{b}}                       
}
\startdata 
CH$_4$ 2.2$\mu$m&\citet{geballe02}   & 0.941$\pm$0.003 & L3.5$\pm$0.5   & 1.047$\pm$0.006 & L6.0$\pm$0.5\\
H$_2$O--2&\citet{slesnick04}\tablenotemark{c}  & 0.799$\pm$0.006 & L2.5$\pm$1.5 & \nodata & \nodata \\
K1&\citet{tokunaga99}\tablenotemark{d} & 0.238$\pm$0.006 & L2.4$\pm$0.9 & 0.21$\pm$0.01 & L1.8$\pm$1.0 \\
H$_2$OD &\citet{mclean03}            & 0.85 $\pm$ 0.07 & L3.3$\pm$1.8 & 0.85$\pm$0.11   & L3.2 $\pm$ 2.9 \\
sH$_2$O$^K$&\citet{testi01}          & 0.27$\pm$0.01   & L2.3$\pm$1.0 & 0.30$\pm$0.05   & L2.9 $\pm$1.0 \\
\enddata
\tablenotetext{a}{Using a Monte Carlo approach to account for uncertainties in the spectra}
\tablenotetext{b}{Using a Monte Carlo approach to account for uncertainties in spectra and the index-SpT relation}
\tablenotetext{c}{valid for spectral types of M2--L3, thus the measurement for \sciencebinb\ is omitted.}
\tablenotetext{d}{Index--SpT relation from \citet{reid01}}
\label{indices}
\end{deluxetable}

%%%%%%%%%%%%%%%%%%%%Table 1%%%%%%%%%%%%
%\begin{deluxetable}{lc}
%\tablecolumns{2}
%\footnotesize
%\tablecaption{Properties of \sciencebin}
%\tablewidth{0pt}
%\tablehead{
%\colhead{}                  &
%\colhead{}             \\
%}
%\startdata
%Angular Separation\tablenotemark{a}                                &  \\
%Position Angle\tablenotemark{a}                         &  \\   
%Distance &  55$\pm$21~pc \\
%Total Mass & \\
%Mass ratio & 0.69$\pm$0.02 \\
%Projected Separation\tablenotemark{b} &  \\
%Orbital Period & xxx$^{+1600}_{-500}$~years \\
%Proper Motion\tablenotemark{c} (mas yr$^{-1}$)             		& 89$\pm$10, 18$\pm$9\\
%$V_{tan}$\tablenotemark{c} (km s$^{-1}$)             		& 21$\pm$5 \\
%\enddata
%\tablenotetext{a}{Values measured from the $J$-band image.}
%\tablenotetext{b}{}
%\tablenotetext{c}{from \citet{scholz09}}
%\label{binprops}
%\end{deluxetable}

%%%%%%%%%%%%%%%%Figures%%%%%%%%%%%%%%%%%

\end{document}